\documentstyle[12pt,epsf]{article}
\renewcommand{\theequation}{\thesection.\arabic{equation}}
\renewcommand{\thefootnote}{\fnsymbol{footnote}}
\newlength{\extraspace}
\newlength{\extraspaces}
\newcommand{\be}{\begin{equation}
\addtolength{\abovedisplayskip}{\extraspaces}
\addtolength{\belowdisplayskip}{\extraspaces}
\addtolength{\abovedisplayshortskip}{\extraspace}
\addtolength{\belowdisplayshortskip}{\extraspace}}
\newcommand{\ee}{\end{equation}}
 
\newcommand{\ba}{\begin{eqnarray}
\addtolength{\abovedisplayskip}{\extraspaces}
\addtolength{\belowdisplayskip}{\extraspaces}
\addtolength{\abovedisplayshortskip}{\extraspace}
\addtolength{\belowdisplayshortskip}{\extraspace}}
\newcommand{\ea}{\end{eqnarray}}

\newcommand{\bas}{\begin{eqnarray*}
\addtolength{\abovedisplayskip}{\extraspaces}
\addtolength{\belowdisplayskip}{\extraspaces}
\addtolength{\abovedisplayshortskip}{\extraspace}
\addtolength{\belowdisplayshortskip}{\extraspace}}
\newcommand{\eas}{\end{eqnarray*}}
 
\newcounter{subequation}[equation]
\makeatletter

\expandafter\let\expandafter
\reset@font\csname reset@font\endcsname

\def\subeqnarray{\arraycolsep1pt
    \def\@eqnnum\stepcounter##1{\stepcounter{subequation}%
        {\reset@font\rm(\theequation\alph{subequation})}}
\jot5mm     \eqnarray}

\makeatother
\newcommand{\newsection}[1]{
\vspace{15mm}
\pagebreak[3]
\addtocounter{section}{1}
\setcounter{equation}{0}
\setcounter{subsection}{0}
\setcounter{footnote}{0}
\begin{flushleft}
{\large\bf \thesection. #1}
\end{flushleft}
\nopagebreak
\medskip
\nopagebreak}
 
\newcommand{\newsubsection}[1]{
\vspace{1cm}
\pagebreak[3]
 
\addtocounter{subsection}{1}
\noindent{ \bf \thesection.\arabic{subsection} #1}
\nopagebreak
\vspace{2mm}
\nopagebreak}

\newcommand{\NP}[1]{Nucl.\ Phys.\ {\bf #1}}
\newcommand{\PL}[1]{Phys.\ Lett.\ {\bf #1}}
\newcommand{\CMP}[1]{Comm.\ Math.\ Phys.\ {\bf #1}}
\newcommand{\PR}[1]{Phys.\ Rev.\ {\bf #1}}
\newcommand{\PRL}[1]{Phys.\ Rev.\ Lett.\ {\bf #1}}

\newcommand{\IJMP}[1]{Int.\ J.\ Mod.\ Phys.\ {\bf #1}}

\newcommand{\R}{\mbox{\rm I\hspace{-.4ex}R}}

\newcommand{\bra}{\langle}
\newcommand{\ket}{\rangle}
\newcommand{\ra}{\rightarrow}

\newcommand{\rra}{\ \longrightarrow \ }

\newcommand{\is}{ &\!=\!& }
\newcommand{\nonum}{\nonumber \\[1.5mm]}
\newcommand{\sspace}{\makebox[1cm]{ }}
\newcommand{\bspace}{\makebox[2cm]{ }}

\newcommand{\lsim}{\raisebox{-.6ex}{${\textstyle\stackrel{<}{\sim}}$}}

\newcommand{\th}{{\theta}}
\newcommand{\lb}{\lambda}
\newcommand{\sh}{{\rm sh}}
\newcommand{\ch}{{\rm ch}}
\renewcommand{\d}{{\partial}}

\newcommand{\cF}{{\cal F}}

\newcommand{\cL}{{\cal L}}

\newcommand{\cO}{{\cal O}}

\begin{document}
%
%%%%%%%%%%%%%%%%%%%%%%%%%%%%%%%%%%%%%%%%%%%%%%%%%%%%%%%%
\begin{titlepage}
%
%footnotesymbols others than numbers
\renewcommand{\thefootnote}{\fnsymbol{footnote}}
\begin{flushright}
%{\bf DRAFT 4}\\
MPI-PhT/96-120\\
hep-th/9612039
\end{flushright}
\vspace{1cm}

\begin{center}
{\LARGE Off-Shell Dynamics of the O$(3)$ NLS Model}\\[3mm]
{\LARGE Beyond Monte-Carlo and Perturbation Theory}
\vspace{1.6cm}
 
{{\large J. Balog\footnote{on leave of absence from the Research 
Institute for Particle and Nuclear Physics, Budapest, Hungary}
\/ and \/ M. Niedermaier}}\\ [3mm]
{\small\sl Max-Planck-Institut f\"{u}r Physik}\\
{(\small\sl Werner Heisenberg Institut)}\\
{\small\sl F\"{o}hringer Ring 6, 80805 Munich, Germany}

\vspace{2.4cm}
{\bf Abstract}
\end{center}

\begin{quote}
The off-shell dynamics of the O$(3)$ nonlinear sigma--model is probed 
in terms of spectral densities and two-point functions by means of the 
form factor approach. The exact form factors of the Spin field, 
Noether--current, EM--tensor and the topological charge density are 
computed up to 6-particles. The corresponding $n\leq 6$ particle 
spectral densities are used to compute the two-point functions, and
are argued to deviate at most a few per mille from the exact answer in 
the entire energy range below $10^3$ in units of the mass gap. To cover 
yet higher energies we propose an extrapolation scheme to arbitrary 
particle numbers based on a novel scaling hypothesis for the spectral 
densities. It yields candidate results for the exact two-point functions 
at all energy scales and allows us to exactly determine the values of 
two, previously unknown, non-perturbative constants. 
\end{quote}
\vfill
\setcounter{footnote}{0}
\end{titlepage}
%%%%%%%%%%%%%%%%%%%%%%%%%%%%%%%%%%%%%%%%%%%%%%%%%%%%%%%%
\renewcommand{\thefootnote}{\arabic{footnote}}
\newsection{Introduction}

The O$(N)$ nonlinear sigma (NLS) models describe the dynamics 
of `Spin' fields $S=(S_1,\ldots,S_N)$ taking values in the $N-1$ 
dimensional unit sphere and governed by the action 
\be
{\cal S} = \frac{1}{2g_0^2}\int d^2 x\,\d_{\mu}S\cdot \d^{\mu}S \;,
\sspace S \cdot S =1\;,
\ee
where $g_0$ is a dimensionless coupling constant. Classically it has 
two important symmetries: First the invariance under the action of 
the internal O$(N)$ group rotating the spins, and second the 
invariance under spacetime conformal transformations. The QFT is
thought to describe an O$(N)$-multiplet of stable massive particles,
the mass scale being non-perturbative in 
the coupling constant \cite{REN,PTasy}.    
Thus, the first of the above classical symmetries is preserved,
but the second is lost. 

It may be worthwhile to briefly recapitulate the physical picture 
behind this phenomenon, which is most succinctly done in the lattice 
formulation. The action (1.1) is replaced by a discretized version on 
some finite $L\times L$ square lattice. The functional integral is then 
well-defined and can be approximately evaluated by means of Monte Carlo 
simulations. The continuum limit is taken by driving the system into a 
critical point. For the bare coupling constant 
it is generally believed that the only critical point is
$g_0 \ra 0$. (See however \cite{PS1,PS2} for the possibility of a 
Kosterlitz-Thouless (KT) type phase transition for $g_0 >0$.) There exist 
non-trivial spin configurations whose energy goes to zero as 
$L \ra \infty$ for fixed $g_0$ and which therefore in infinite volume 
are present at arbitrarily 
small $g_0$. These so-called super-instanton configurations can be 
thought of being the ``enforcers'' of the Mermin-Wagner theorem \cite{PS1,PS2}.
They disorder the spins, forbidding a spontaneous magnetization in two 
dimensions. In QFT language this means that the O$(N)$ symmetry is unbroken
and (in the absence of a KT-phase transition) the theory has a mass gap. 
It also means that perturbation theory (PT),
known to be renormalizable in finite volume \cite{REN}, has infrared
problems as it starts from a fictitious ordered ground state. 
Nevertheless for O($N$) invariant correlation functions a result by 
Elitzur and David \cite{D} guarantees that (using periodic or Dirichlet 
boundary conditions) the coefficients of the bare lattice PT expansion 
have finite $L\ra \infty$ limits. The bare PT expansion can then be 
converted into a renormalized one, and the renormalized one into an 
expansion in the running coupling constant, where the running is defined 
through the perturbative beta function \cite{PTasy}. It must be emphasized 
that all this can be (and is) done regardless whether or not the final 
expansion coefficients
have any relation to, and significance for, the unknown exact correlation
functions. The hypothesis of asymptotic freedom is that they do have.
Namely, if one were given the exact correlation functions and tried to 
perform an asymptotic expansion in the running coupling constant
(the running again defined by the perturbative beta function) the claim is:
(i) this expansion exists and (ii) the coefficients obtained coincide with
the ones computed in PT.   
   
This is a mathematically precise statement which is either true or false.
It is true that it is commonly believed to be true -- though not everyone
might wish to take this as a substitute for a proof. Indeed, the correctness
of the claims (i) and (ii) has been challenged in a series of papers
\cite{PS1,PS2}, stimulating some controversy \cite{Contr}. Evidently the 
problem cannot be addressed within the perturbative frame. It is also 
difficult to even come close to conclusive results on the basis of Monte Carlo 
simulations alone; the presently available data cover only the energy 
range below $50m$, where $m$ is the mass gap of the model. This is far too 
low to study the elusive high energy properties. The purpose of this and
an accompanying technical paper \cite{BN2} is to study the off-shell
dynamics of the O$(3)$ NLS model via the form factor approach.      
Form factors in this context are the matrix elements of some local 
operator between the physical vacuum and some asymptotic multi-particle 
states. In a QFT with a factorized scattering theory they can in 
principle be determined exactly by solving a set of recursive functional
equations \cite{Weiszetal,Smir} 
that use the exact two-particle S-matrix as an input.
Once the form factors are known, off-shell quantities, e.g.~two-point
functions, can be computed by inserting a resolution of the 
identity in terms of multi-particle states. For models whose S-matrix is
diagonal in isospin space, the resulting low energy expansion has been 
seen to converge  
rather rapidly \cite{YZ,Conv1,Conv2}. Models with a non-diagonal S-matrix are 
technically much more demanding. However they also provide a much better 
testing ground for 4-dimensional QFT scenarios and deserve more dedicated 
attempts to gain insight into their exact off-shell dynamics. 
  
In the O$(3)$ NLS model we computed the exact form factors of the 
Spin field, Noether--current, EM--tensor and the topological charge 
density up to 6-particles. Their two-point functions are
evaluated by means of a K\"allen-Lehmann spectral representation 
\cite{KLehmann}. The $n\leq 6$ particle spectral densities are argued to 
provide results for the two-point functions that differ only about 3 to 
10 per mille from the exact answer in the entire energy range 
below $10^3m-10^4m$, depending on the quantity considered \cite{BHN}. 
In the low energy range the agreement with the Monte Carlo data
of Patrascioiu and Seiler \cite{PSMC} is excellent. At higher energies we 
compare with renormalization group improved 2-loop PT. Within the 
range considered 2-loop PT yields an (within 1\%) accurate
description of the system only for energies above $50m-100m$, provided one 
uses the known exact value of the Lambda parameter \cite{HN} to 
fix its absolute normalization. Based on the low energy Monte Carlo 
data alone, however, one would be tempted to maximize the apparent 
domain of validity of PT by tuning the Lambda parameter such as to 
match the relevant part of the Monte Carlo data. Doing this in the 
NLS model the Lambda parameter comes out wrong by about 10\%. Generally 
speaking this emphasizes the importance to have an 
independent estimate for the onset of the (2-loop) perturbative 
regime. In the case at hand this is provided by the form factor
results.

A major challenge remains the computation of the extreme UV 
properties of the model {\em independent} of PT. For the important
case of a two-point function this amounts to summing 
up all multi-particle contributions to their spectral resolution. 
The short-distance asymptotics of the two-point functions is related
to the $\mu \ra \infty$ asymptotics of the corresponding spectral
density $\rho(\mu)$. We approach the problem by taking advantage
of a remarkable self-similarity property of the $n$-particle
spectral densities: For large $n$ and $\lb >1$ they appear to 
behave like 
\be
\rho^{(\lb n)}(\mu) \approx \frac{m}{\mu \lb^{\gamma}}\,
(\mu/m)^{1/{\lb^{1+\alpha}}} \;
\rho^{(n)}\left(m(\mu/m)^{1/{\lb^{1+\alpha}}}\right)\;,
\ee
where $m$ is the mass gap and $\gamma,\alpha$ are certain 
`critical' exponents. Given $\rho^{(n)}(\mu)$ for some initial 
particle number $n$, (1.2) can be used  -- without actually
doing the computation (which for large $N$ is practically 
impossible) -- to anticipate the structure of all $N >n$ 
particle spectral densities. We promoted (1.2) to a working
hypothesis and explored its consequences. The results obtained 
and the status of the hypothesis may be surveyed as follows:
\begin{itemize}
\item It is a statement about the spectral densities of 
the exact theory at {\em all} energy scales. 
\item Its consequences for the UV behavior are consistent with 
PT.
\item It produces new exact non-perturbative results like 
the normalization of the spin two-point function at short distances.
\item It allows one to compute numerically the two-point functions 
at {\em all} energy/length scales.
\end{itemize}
The paper is organized in the following way. In section 2 we 
describe general features of the spectral representation of 
two-point functions and their asymptotic expansions. Applied to the 
O$(3)$ NLS model we prepare various results on the four operators
under consideration. The next section contains a summary of  
our results on the O$(3)$ form factors, form factor squares and their 
properties, as well as an exact expression for the asymptotics of the
$n$-particle spectral densities. The results on the two-point functions 
based on the $n\leq 6$ particle form factors are described in section 4. 
The extrapolation to arbitrary particle numbers by means of the  
above scaling hypothesis is implemented in section 5, to be 
followed by brief conclusions.

%%%%%%%%%%%%%%%%%%%%%%%%%%%%%%%%%%%%%%%%%%%%%%%%%%%%%%%%%%%%%%%%%%
%\include{sect2}
%\newpage
\newsection{Spectral representation of two-point functions}
\vspace{-1cm}

\newsubsection{Spectral representation} 

The form factors characterize an (integrable as well as non-integrable)
QFT in a similar way as the $n$-point functions do. Assuming
the existence of a resolution of the identity in terms of asymptotic
multi-particle states, the $n$-point functions can in principle be 
recovered from the form factors. In the important case of the two-point
functions this amounts to the well-known spectral representation. 
For the Minkowski two-point function (Wightman function) 
of some local operator $\cO$ one obtains in a first step
\be
W^{\cO}(x-y)= 
\sum_{n\geq 1}\frac{1}{n!}\int \prod_{j=1}^n \frac{d\th_j}{4\pi}\;
e^{-i(x^0-y^0)P_0^{(n)}(\th) - i(x^1 -y^1)P_1^{(n)}(\th)}\;
|\cF^{(n)}(\th)|^2 \;,
\label{sp1}
\ee  
where $\cF^{(n)}(\th) = \bra 0 |\cO(0)|\th_n,\ldots,\th_1\ket$ are the 
form factors of $\cO$ and $P_{\mu}^{(n)}(\th) = \sum_i p_{\mu}(\th_i)$,
with $p_0(\th) = m\ch\th,\;p_1(\th) = m\sh \th$ are the eigenvalues of 
energy and momentum on an $n$-particle state.% 
\footnote{Our kinematical conventions are: $(x^0,x^1)$ are coordinates
on 2-dimensional Minkowski space $\R^{1,1}$ with bilinear form  
$x\cdot y = x^{\mu}\eta_{\mu\nu}y^{\nu},\;\eta =\mbox{diag}(1,-1)$. 
Lightcone coordinates are $x^{\pm} = (x^0\pm x^1)/\sqrt{2}=x_{\mp}$; 
the norm is $\|x\| = \sqrt{\pm x\cdot x}= \sqrt{\pm2 x^+x^-},\;\pm 
x^2\geq 0$. The antisymmetric tensor is $\epsilon_{01}=-\epsilon_{10}=1$. 
The normalization of the 1-particle states is 
$\bra\th_1|\th_2\ket =4\pi\,\delta(\th_1-\th_2)$, which corresponds to
the standard normalization in $d+1$ dimensions, specialized to $d=1$.
For simplicity we assume that all particles are of the same mass $m$ and 
suppress internal indices in section 2.1.}
The local operators are classified by various quantum numbers, in 
particular by their Lorentz spin $s$ and their mass dimension $\Delta$. 
It turns out to be convenient to parametrize their form factors in 
terms of `scalarized form factors', which are functions
of the rapidity differences only and carry quantum numbers $\Delta = s =0$.
For an operator with quantum numbers $(\Delta, s)$ we shall use the 
parametrization 
\be
\cF^{(n)}(\th) = \cL(P^{(n)}(\th)) \,f^{(n)}(\th)\;,
\label{sp2}
\ee
where in the cases of interest $\cL$ is a polynomial 
of degree $\Delta \geq |s|$ in the $n$-particle momenta  
$P^{(n)}(\th)$ of integer spin $s$, i.e.
\be
\cL(P^{(n)}(\th)) = e^{-su} \cL(P^{(n)}(\th + u)) \;.
\label{sp3}
\ee
The Wightman function (\ref{sp1}), considered as a distribution, can then
be obtained by differentiation
\be
W^{\cO}(x-y) = \cL(i\d_x)\,\cL(i\d_y)\, W(x-y)\;,
\label{sp4}
\ee
where $W(x-y)$ is defined as the r.h.s.~of (\ref{sp1})  with 
$|\cF^{(n)}(\th)|^2$ replaced by $|f^{(n)}(\th)|^2$. We shall refer to
$W(x)$ as the ``scalarized Wightman function of $\cO$".  
Let us emphasize that in general $W(x-y)$ can {\em not} be interpreted
as the two-point function of some local (scalar) field; it is only
a useful auxiliary function from which the physical 2-point function can
be obtained through differentiation. 
  
For many purposes it is useful to rewrite (\ref{sp1}) in terms of a 
K\"allen-Lehmann spectral representation \cite{KLehmann}. 
Changing integration variables 
according to 
\ba
&& u_i = \th_i -\th_{i+1}\;,\sspace 1\leq i\leq n-1\;,\sspace
\alpha = \ln\left(
\frac{m(e^{\th_1}+\ldots + e^{\th_n})}{M^{(n)}(u)} \right)\;,\nonum  
&& M^{(n)}(u) = m
\left[n + 2 \sum_{i<j}\ch(u_i+\ldots +u_{j-1})\right]^{1/2}
\label{sp5}
\ea    
one obtains for the scalarized Wightman function
\ba
&& W(x-y) = -i\int_0^{\infty}d\mu\;\rho(\mu)\;
D(x-y;\mu)\;,\nonum
&& \rho(\mu) = \sum_{n\geq 1}\rho^{(n)}(\mu)\;,\sspace 
\rho^{(1)}(\mu) = \delta(\mu - m)\,|f^{(1)}|^2\;,\nonum
&& \rho^{(n)}(\mu) = \int_0^{\infty}
\frac{d u_1 \ldots d u_{n-1}}{(4 \pi)^{n-1}}\;|f^{(n)}(u)|^2\; 
\delta(\mu - M^{(n)}(u))\;,\;\;\;n\geq 2\;.
\label{sp6}
\ea
Notice that no problem of convergence arises for the spectral density.
First, each $n$-particle contribution exists because, 
for fixed $\mu$, the integrand has support only in a compact domain 
$V(\mu)\subset\R_+^{n-1}$ in which the form factors are bounded functions
so that the integration is well-defined.
Viewed as a function of $\mu$ one observes that $\rho^{(n)}(\mu)$ has 
support only for $\mu \geq m n$. Summing up the $n$-particle contributions 
to $\rho(\mu)$ therefore only a finite number of terms (those with 
$n\leq [\mu/m]$, $[x]$ being the integer part of $x$) contribute.  
Under some mild assumptions on the growth of $\rho(\mu)$ 
(for example it is sufficient to require that $\rho(\mu)$ is
polynomially bounded in $\mu$)
the existence 
of the spectral density guarantees that of the 2-point function, as
defined through (\ref{sp6}). The integration kernel is given by
\be
D(x;m) = \frac{1}{4}\,\th(x^2)
\left[ \mbox{sign}(x_0)\,J_0(m\|x\|) - i Y_0(m\|x\|) \right] + 
\frac{i}{2\pi}\th(-x^2) K_0(m\|x\|)\;,
\label{sp7}
\ee 
and coincides with the 2-point function of a free scalar field of 
mass $m$. We use the conventions of \cite{MOS} for the Bessel functions. 
The differentiation (\ref{sp4}) is a bit cumbersome in the general case;
usually however one will be interested in the behavior at spacelike 
distances in which case only higher order modified Bessel function 
$K_{n}(m\|x\|)$ arise. Alternatively one can use the Fourier
representation of $D(x;m)$ and write
\ba
W^{\cO}(x-y) \is \int_0^\infty d\mu \rho(\mu)
\int\frac{d^2 p}{2\pi}\,\theta(p_0)
\delta(p^2 - \mu^2)\;\cL(p)\,\cL(-p)\;e^{-ip\cdot (x-y)}\;. 
\label{sp8}
\ea

On general grounds the Fourier transform of $W^{\cO}(x)$ must have 
support only inside the forward lightcone 
$V^+ = \{p\in \R^{1,1}\,|\, p^2 >0,\,p_0>0\}$. 
From (\ref{sp8}) one finds indeed
\bas
&& \widetilde{W}^{\cO}(p) = \left\{ \begin{array}{cl}
{\displaystyle \frac{\pi \rho(\|p\|)}{\|p\|}}\,
\cL(p)\cL(-p) & \sspace{ p\in V^+} \\ 
0 & \sspace{ p\notin V^+} \end{array}\right.
\eas
Thus, up to kinematical factors the spectral density can also 
be viewed as the Fourier transform of the two-point Wightman 
function. 

For comparison with perturbation theory one needs the 
time-ordered two-point function and its Fourier transform. Its 
spectral representation is easily read off from (\ref{sp6}), (\ref{sp8}) 
\be
G^{\cO}(x-y) = -i\int\frac{d^2 p}{(2\pi)^2}\,e^{-ip\cdot (x-y)}\;
\cL(p)\,\cL(-p)\; I(-p^2 - i\epsilon)\,,
\label{sp9}
\ee
where
\be
I(z) = \int_0^{\infty}d\mu\frac{\rho(\mu)}{z+\mu^2}\;.
\label{sp10}
\ee
The definition of $I(z)$ has been chosen such that it has a cut along the
negative real axis and one can recover the spectral density from the
discontinuity along this cut 
$$
\rho(\mu)=\frac{i\mu}{\pi}\,{\rm disc}\,I(-\mu^2)\,.
$$
Conversely, ${\rm disc}\,I(z)$ determines $I(z)$ up to a polynomial 
ambiguity. 

The spectral representation of the Euclidean two-point
function (Schwinger function) is obtained similarly. The Schwinger 
function can be defined by $S^{\cO}(x_1,x_2) := W^{\cO}(-ix_2,x_1)$ 
for $x_2>0$ and then by analytic continuation to $x_2<0$. By 
construction it also coincides with the analytic continuation of 
$G^{\cO}(x)$. In a momentum space integral this formally amounts
to the replacement $(p_0,p_1) = (ip^E_2,p_1^E)$. The Euclidean 
counterparts of (\ref{sp8}) and (\ref{sp9}) thus are
\ba
S^{\cO}(x-y) \is  \int\frac{d^2 p}{(2\pi)^2}\,e^{-ip\cdot (x-y)}\;
\cL^E(p)\,\cL^E(-p)\; I(p^2)\nonum
\is \cL^E(i\partial_x) \cL^E(i\partial_y) \,S(x-y)\;. 
\label{sp11}
\ea
We shall also use the notation $\bra \cO(x)\cO(y)\ket $ for $S^{\cO}(x-y)$.
The integration in (\ref{sp11}) is now over the Euclidean momenta 
$(p_1,p_2) = (p_1^E, p_2^E)$ and $\cL^E(p_1,p_2) := \cL(ip_2, p_1)$. The 
``scalarized Schwinger function of $\cO$" entering the second line is 
\be
S(x)=\frac{1}{2\pi}\,\int_0^\infty d\mu\,\rho(\mu)\, 
K_0\left(\mu\sqrt{x_1^2 + x_2^2}\right)\,.
\label{sp12}
\ee
Notice that the spectral density $\rho(\mu)$ and hence the function 
$I(z)$ is the same in the Minkowski and in the Euclidean situation. 
Given either $\rho(\mu)$ or $I(z)$ for a specific operator all two-point 
functions can be computed -- scalarized and physical, Minkowski and
Euclidean ones.  

As an example consider the case of the energy momentum (EM) tensor. 
It is of additional interest, because its spectral density is closely
related to the Zamolodchikov C-function. In $1+1$ dimensions, the symmetric
and conserved EM tensor can always be parametrized in terms of a
non-local scalar field $\tau$, the \lq\lq EM-potential"
\be
T_{\mu\nu}(x)=\cL_{\mu\nu}(i\d_x)\,\tau(x)\,,\;\sspace
\cL_{\mu\nu}(p) :=- p_{\mu}p_{\nu}+\eta_{\mu\nu}p^2\,.
\label{EM1}
\ee
The defining relation (\ref{sp2}) for the scalarized EM form
factors thus reads 
\be
\bra 0|T_{\mu\nu}(0)|\th_n,\ldots,\th_1\ket = 
\left[ - P^{(n)}_{\mu}(\th)  P^{(n)}_{\nu}(\th) +\eta_{\mu\nu}\, 
 P^{(n)}_{\rho}(\th)  {P^{(n)}}^{\rho}(\th) \right] \;
f^{(n)}(\th_n\ldots \th_1)\;, 
\label{EM2}
\ee
and the scalarized form factors can be interpreted as that  of the scalar 
field $\tau$. Using (\ref{sp6}) the spectral representation of 
the Minkowski two-point function is 
\be
\bra0|T_{\mu\nu}(x) T_{\alpha\beta}(y)|0\ket = 
-i\cL_{\mu\nu}(i\d_x)\cL_{\alpha\beta}(i\d_y) 
\int_0^{\infty}d\mu\;\rho(\mu)\;D(x-y;\mu)\,,
\label{EM3}
\ee
and similarly for the time-ordered two-point function. Their analytic
continuations to $x_0 = -ix_2$ coincide and yield the Schwinger function
according to the rules (\ref{sp11}), (\ref{sp12}) with 
$\cL^E_{\alpha\beta}(p) = -p_{\alpha}p_{\beta} +\delta_{\alpha\beta}p^2$.
Explicitly
\ba
\bra T_{\mu\nu}(x)T_{\alpha\beta}(y)\ket &=& \int\,\frac{d^2p}{(2\pi)^2}
e^{-ip\cdot(x-y)}\,[p_\mu p_\nu-\delta_{\mu\nu}\,p^2]\,
[p_\alpha p_\beta-\delta_{\alpha\beta}\,p^2]\, I(p^2)\nonum
&=&\cL_{\mu\nu}^E(i\d_x)\cL^E_{\alpha\beta}(i\d_y)\,S(x-y)\,.
\label{EMEu}
\ea
The lightcone components correspond to the $SO(1,1)$-irreducible 
pieces. In particular $T_{++}$ and the trace $t=2T_{+-}$ 
have polynomials $\cL_{++}(p) = -p_+^2$ and $2\cL_{+-} = p^2$
respectively. Combining (\ref{sp7}) 
and (\ref{sp6}) one finds for the behavior at small spacelike distances
\be
\bra0|T_{++}(x)T_{++}(0)|0\ket = \frac{c}{2\pi^2}
\left(\frac{x^-}{x^+}\right)^2\;\frac{1}{(x^2)^2}+\ldots\,, 
\sspace x^2 <0\,,\;\;\; x^2\to0\,,
\label{EM4}
\ee
where the dots stand for less singular terms whose form is 
model-dependent. The coefficient $c$ is given by 
\be
c = 12 \pi \int_0^{\infty}d\mu \,\rho(\mu)
\label{cdef}
\ee
and coincides with the central charge of the Virasoro algebra in the 
conformal field theory describing the UV  fixed point of the model. 
This latter fact is part of the statement of Zamolodchikov's 
C-theorem \cite{ZC}.%
\footnote{In particular c is finite; for other operators the integral 
over the spectral density will in general not converge.}
The proof of the C-theorem is particularly transparent from the viewpoint 
of the spectral representation in the Euclidean case \cite{CFL}. 
In particular using (\ref{EMEu}), (\ref{sp12}), the properties of 
the modified Bessel functions and the definition (\ref{cdef}) 
it is easy to arrive at the ``C-theorem sum rule" \cite{Csum} for 
theories with a mass gap
$$
c=3\pi\,\int\, d^2x\, x^2\;\bra t(x)\,t(0)\ket\,,
\label{Cth}
$$
where $\bra t(x)\,t(0)\ket$ is the Euclidean correlator of the trace. 

%%%%%%%%%%%%%%%%%%%%%%%%%%%%%%%%%%%%%%%%%%%%%%%%%%%%%%%%%%%%%%%%%%%%%%%%
\newsubsection{Asymptotic expansions}

Suppose that in a QFT the validity of asymptotically free perturbation 
theory has been established in the sense outlined in the introduction. 
Then PT allows one to compute the asymptotic expansion of the Fourier 
transform of the two-point functions for large Euclidean momenta. This 
can be translated into the language of the spectral density using 
the defining equation (\ref{sp10}). Motivated by the asymptotically free 
case, we define $\iota(\xi)$ by
\be
I(z)=\frac{1}{z}\,\iota\,\left({\rm ln}\,\frac{\sqrt{z}}{m}\right)
\label{iotadef}
\ee
and $R(\xi)$ by
\be
\rho(\mu)=\frac{1}{\mu}\,R\,\left({\rm ln}\,\frac{\mu}{m}\right)\,.
\label{Rdef}
\ee
Using the relation (\ref{sp10})
we can compute the asymptotic expansion of $R(\xi)$ in terms of 
$\iota(\xi)$ and its derivatives:
\be
R(\xi)\,\dot{\approx}\,\sum_{s=0}^\infty H_{2s+1}\,
\iota^{(2s+1)}(\xi),
\label{AS1}
\ee
where $H_1=1$, $H_3=-\frac{1}{4}\zeta(2)$ and all the numerical coefficients
$H_{2s+1}$ are  polynomials in
\be
\zeta(k)=\sum_{n=1}^\infty\,\frac{1}{n^k}\sspace k\geq 2\,.
\label{AS2}
\ee
In (\ref{AS1}) the symbol $\,\dot{\approx}\,$ signifies 
that both sides of the equation have the same asymptotic expansion in 
$1/\xi$, i.e. their difference is smaller than any power of $1/\xi$
for large $\xi$. 
In addition to the asymptotic relation (\ref{AS1}), there is a relation 
between the integral of $R(\xi)$ and the leading term in the asymptotic 
expansion of $\iota(\xi)$ (provided, of course, both are finite):
\be
\iota(\infty)=\int_0^\infty\, d\xi\, R(\xi)=\int_0^\infty\, 
d\mu\,\rho(\mu)\,.
\label{AS3}
\ee
In perturbation theory the asymptotic expansion is usually given with
the help of some running coupling constant, in terms of which the expansion is
a power series and which goes to zero as the momentum goes to infinity.
There are many functions having these properties. A simple and convenient 
choice is $e(\xi)$ defined by
\be
\frac{1}{e(\xi)}+\kappa\,{\rm ln}\, e(\xi)=\xi\,,
\label{AS4}
\ee
where the parameter $\kappa$ is related to the first two beta function 
coefficients of the model. We shall refer to $e(\xi)$ as the universal 
running coupling function because only the universal part of the beta 
function enters its definition. Any other running coupling function can 
then be expressed as a power series in $e(\xi)$. In particular this 
would hold for the lattice-measurable running coupling (for various possible 
definitions of the latter, see \cite{running}). 

If the expansion of the function $\iota$ starts off at
the $p^{\rm th}$ power of $e$
\be
\iota(\xi)\;\dot{\approx}\;\iota_0\, 
e^p+\iota_1\, e^{p+1}+\iota_2\, e^{p+2}+\cdots
\label{AS5}
\ee
using (\ref{AS1}) we can write
\be
R(\xi)\;\dot{\approx}\; 
r_0\, e^{p+1}+r_1\, e^{p+2}+r_2\, e^{p+3}+\cdots\,,
\label{AS6}
\ee
where
\ba
r_0\is -p\,\iota_0\,,\cr
r_1\is -p\kappa\,\iota_0-(p+1)\iota_1\,,\cr
r_2\is \frac{1}{4}\iota_0\,\zeta(2)\,p\,(p+1)\,(p+2)-p\kappa^2\,\iota_0-
(p+1)\,\kappa\,\iota_1-(p+2)\,\iota_2\,.
\label{AS7}
\ea

%%%%%%%%%%%%%%%%%%%%%%%%%%%%%%%%%%%%%%%%%%%%%%%%%%%%%%%%%%%%%%%%%%%%%
\newsubsection{O($3$) Nonlinear Sigma-Model}
   
The O$(3)$ non-linear $\sigma$-model is asymptotically free in perturbation
theory and admits an exact on-shell solution via the the bootstrap method. 
The asymptotic single particle states are assumed to consist of an O$(3)$ 
triplet of massive particles with mass $m$: $|a,\th\ket$, $a=1,2,3$. In 
this paper we will study the properties of the four most interesting local 
operators in the O$(3)$ model: The spin field $\Phi^a$, the Noether current
$J^a_\mu$, the EM tensor $T_{\mu\nu}$ and the topological change density $q$.
In the following we shall discuss these four operators consecutively.

{\em Spin field:} We normalize our spin field operator $\Phi^a(x)$ by
\be
\bra 0|\Phi^a(0)|b,\th\ket=\delta^{ab}\,.
\label{o31}
\ee
$\Phi^a$ is proportional to the Lagrangian field renormalized in (say) the 
$\overline{\rm MS}$ scheme
\be
\Phi^a(x)=\zeta\,S^a_{{\overline {\rm MS}}}(x)\,,
\label{o32}
\ee
where $\zeta$ is some unknown finite renormalization constant.
The $n$-particle form factor of $\Phi^a$ is defined by
\be
\bra 0|\Phi^a(0)|a_n,\th_n;\dots;a_1,\th_1\ket=\frac{2}{\sqrt{\pi}}\,
f^a_{1,a_n\cdots a_1}(
\th_n,\dots,\th_1)\,,\sspace (\mbox{$n$ odd})\;.
\label{o33}
\ee
Here we introduced the normalization factor $2/\sqrt{\pi}$ for
later convenience (c.f. section 3.2). The subscript `1' on the 
r.h.s.~indicates that under an O$(3)$ rotation these functions 
are intertwiners ${\bf 3}^{\otimes n} \ra {\bf 3}$ onto the 
irreducible representation ${\bf 3}$ of isospin $1$. 
Only the odd particle form factors are non-vanishing because $\Phi^a$ 
is odd under the internal parity reflection of the O$(3)$ variable. 
In addition they depend only on the rapidity differences. As a 
consequence one can write
\be
\delta^{ab}\, F^{(n)}_1(u) :=\sum_{a_1\dots a_n}
f^a_{1,a_n\cdots a_1}(\th_n,\dots,\th_1)^*
f^b_{1,a_n\cdots a_1}(\th_n,\dots,\th_1)\,.
\label{o34}
\ee
The spectral density is given by
\begin{subeqnarray}
\rho^{\rm spin}(\mu) \is \frac{4}{\pi}
\sum_{k=0}^\infty\rho^{(2k+1)}_1(\mu)\,,\\
\rho^{(n)}_1(\mu) \is \int_0^{\infty}
\frac{d u_1 \ldots d u_{n-1}}{(4 \pi)^{n-1}}\;F_1^{(n)}(u)\; 
\delta(\mu - M^{(n)}(u))\;.
\label{o35}
\end{subeqnarray}
From $\rho^{\rm spin}(\mu)$ or its Stieltjes transform 
\be
I^{\rm spin}(z) = \int_0^{\infty}d\mu\,\,\frac{\rho^{\rm spin}(\mu)}{z+\mu^2}\;
\label{o36}
\ee
all two-point functions of the Spin field can be computed. 

For example the Schwinger function is 
\ba
\bra \Phi^a(x)\,\Phi^b(y)\ket
\is \delta^{ab}\, S^{\rm spin}(x-y)\cr
\is \delta^{ab}\,\int\frac{d^2 p}{(2\pi)^2}\, e^{-ip\cdot(x-y)}\,
I^{\rm spin}(p^2)\,.
\label{o37}
\ea
In the time-ordered Minkowski two-point function 
$-iI^{\rm spin}(-p^2 -i\epsilon)$ would enter. The normalization (\ref{o31}) 
is thus equivalent to the usual (re-)normalization condition on the Fourier
transform of the time ordered Minkowski two-point function
\be
-i\,I^{\rm spin}(-p^2-i\epsilon) = \frac{i}{p^2-m^2+i\epsilon} +\ldots\;,
\;\;\; p^2\approx m^2\;.
\label{o38}
\ee
It also implies that for large (spacelike) distances all two-point 
functions decay like in the free theory. In particular for the Schwinger
function this means 
\be
S^{\rm spin}(x) = \frac{1}{\sqrt{8\pi\, m r}}\,
e^{-m r}\left[ 1 + O(1/ r)\right]\;,\sspace r= \sqrt{x_1^2+x_2^2}\;.
\label{o39}
\ee
We will later determine the spin-spin spectral density and
two-point function by the form factor bootstrap method. We will then be 
able to compare their asymptotic behavior with the results obtained in 
perturbation theory. We shall denote the universal
running coupling function for the $O(3)$ model by $\alpha(p)$, which is 
defined as in the previous subsection to be the solution of
\be
\frac{1}{\alpha(p)}+\ln \alpha(p)=\ln\frac{p}{m}\,.
\label{o310}
\ee
(The $\kappa$ parameter of (\ref{AS4}) is equal to unity for the
O$(3)$ model.) Perturbation theory at 2-loop order predicts for the 
large $p$ asymptotics \cite{UWolff} 
\be
p^2\, I^{\rm spin}(p^2)=\lambda_1\left\{\frac{1}{\alpha(p)}+(2+\xi_0)+
(2+\xi_0)\alpha(p)+O\left(\alpha^2(p)\right)\right\}\,.
\label{o311}
\ee
Here the overall constant $\lambda_1$ is related to the unknown finite
renormalization in (\ref{o32}), while the parameter $\xi_0$ gives the 
connection between the perturbative mass parameter
$\Lambda_{\overline{\rm MS}}$ and the exact mass gap $m$. 
In the O$(3)$ model the latter is known exactly \cite{HN}:
\be
\xi_0=\ln \frac{m}{\Lambda_{\overline{\rm MS}}}=\ln 8-1\,\,\approx\,\,
1.07944\,.
\label{o312}
\ee
Finally, using the results (\ref{AS5}), (\ref{AS6}) and (\ref{AS7})
we obtain the perturbative prediction for the
asymptotic expansion of the spin-spin spectral density
\be
\rho^{\rm spin}(\mu)=\frac{\lambda_1}{\mu}\left\{1+\alpha(\mu)+
O\left(\alpha^2(\mu)\right)\right\}\,.
\label{o313}
\ee
In section 5 we shall find the value of the undetermined non-perturbative
constant $\lb_1$ to be 
\be
\lb_1 = \frac{4}{3\pi^2}\;.
\label{lb1}
\ee
Note that the perturbative series for both the Fourier transform and the
spectral density contain integer powers of the running coupling only.
This fact, which will play an important role in what follows, is a 
special feature of the O$(3)$ model, since for the  otherwise rather
similar O$(N)$ models with $N>3$ one has 
\be
p^2\, I^{\rm spin}(p^2)\sim\left[\alpha(p)\right]^{-\frac{1}{N-2}}\sspace
\rho^{\rm spin}(\mu)\sim\frac{1}{\mu}\left[\alpha(\mu)\right]^{
\frac{N-3}{N-2}}\,.
\label{o314}
\ee
%%%%%%%%%%%%%%%%%%%%%%%

{\em Noether current:} In terms of the Lagrangian variable the Noether 
current is 
\be
J^a_\mu=\frac{1}{g_0^2}\epsilon^{abc}S^b\d_\mu S^c\,.
\label{o315}
\ee
In the QFT we fix the normalization such that its time components satisfy 
the equal time commutation relations
\be
\left[J^a_0(0,x)\;,\;J^b_0(0,y)\right]=i\epsilon^{abc}\, J^c_0(0,y)\,
\delta(x-y)\,.
\label{o316}
\ee
In $1+1$ dimensions a conserved current can always be parametrized in 
terms of a (non-local) scalar field, the \lq\lq current potential" $j^a$
via
\be
J^a_\mu=\epsilon_{\mu\alpha}\,\d^\alpha\,j^a\,.
\label{o317}
\ee
The scalarized form factors of $J^a_{\mu}$ can be identified with that 
of the field $j^a$. Their defining relation is 
\be
\bra 0|J^a_{\mu}(0)|a_n,\th_n;\dots;a_1,\th_1\ket= \cL_{\mu}(P^{(n)}(\th))
\;f^a_{1,a_n\cdots a_1}(\th_n,\dots,\th_1)\,,\sspace (\mbox{$n$ even})
\label{o318}
\ee
where $\cL_{\mu}(p) = -i \epsilon_{\mu\nu}p^{\nu}$. Note that no confusion 
can arise here from using the same symbol $f^a_{1,a_n\cdots a_1}$ as in 
(\ref{o33}) since there it is defined for an odd number of particles only. 
We also note that (\ref{o316}) fixes the residue of the two-particle 
form factor
\be
f^a_{1,a_2a_1}(\th_2,\th_1) = 
\frac{-2\epsilon^{aa_2a_1}}{\th_2-\th_1-i\pi}+ \ldots \;,
\;\;\; \th_2\approx \th_1+i\pi\,.
\label{o319}
\ee

If we now extend the definitions (\ref{o34}), (\ref{o35}b) 
for even values of $n$ as well, the spectral density in the
case of the Noether current is given by
\be
\rho^{\rm curr}(\mu) = \sum_{k=1}^\infty\rho^{(2k)}_1(\mu)\,.
\label{o320}
\ee
Finally, the Fourier transform for the current is defined as in (\ref{o36})
with the operator superscript $^{\rm spin}$ replaced by $^{\rm curr}$.
Using $\rho^{\rm curr}$ and $I^{\rm curr}$ one can represent all 
the two-point functions of the Noether current. For example the
Schwinger function is
\be
\bra J_\mu^a(x)\,J_\nu^b(y)\ket
= \delta^{ab}\,\int\frac{d^2 p}{(2\pi)^2}\, e^{-ip\cdot(x-y)}\,
[p_\mu\, p_\nu \, -p^2\,\delta_{\mu\nu}]\,
I^{\rm curr}(p^2)\,.
\label{o321}
\ee

Perturbation theory predicts for the Fourier transform \cite{B}
\be
p^2\, I^{\rm curr}(p^2)=\frac{1}{3\pi}\left\{\frac{1}{\alpha(p)}+(\xi_0-1)+
(\xi_0-1)\alpha(p)+O\left(\alpha^2(p)\right)\right\}
\label{o322}
\ee
and, using (\ref{AS5}), (\ref{AS6}) and (\ref{AS7}), for the spectral density
\be
\rho^{\rm curr}(\mu)=\frac{1}{3\pi\mu}\left\{1+\alpha(\mu)+
O\left(\alpha^2(\mu)\right)\right\}\,.
\label{o323}
\ee

{\em Topological charge density:} In terms of the Lagrangian variable the
topological charge density reads
\be
q=\frac{1}{8\pi}\epsilon^{abc}\,\epsilon^{\mu\nu}\, 
S^a\,\d_\mu S^b\,\d_\nu S^c\,.
\label{qdef}
\ee
We shall use a different symbol for its Euclidean version
\be
k=\frac{1}{8\pi}\epsilon^{abc}\,\epsilon^{({\rm E})}_{\alpha\beta}\, 
S^a\,\d_\alpha S^b\,\d_\beta S^c\,,
\label{kdef}
\ee
(where $\epsilon^{({\rm E})}_{12}=-\epsilon^{({\rm E})}_{21}=1$) 
to emphasize that it differs by an extra factor $-i$ from the Euclidean 
continuation of $q$: $k=-i\, q^{({\rm E})}$. This extra factor of $-i$, 
which is a consequence of the linearity of $q$ in the time derivative, 
is responsible for the important fact that the Euclidean two-point 
function of $k$ is a strictly negative function for all non-zero 
Euclidean distances.

A result of the form factor approach is that the topological charge 
density operator can be parametrized in terms of a dimensionless 
non-local scalar 
field $\Phi$ as follows (c.f.~section 3)
\be
q=(\Box +m^2)\,\Phi\,.
\label{Phidef}
\ee
The scalarized form factors of $q$ can be identified with that of $\Phi$. 
Their defining relation is 
\be
\bra 0|q(0)|a_n,\th_n;\dots;a_1,\th_1\ket=\sqrt{\lambda_0}\,\cL(P^{(n)}(\th))
\;f_{0,a_n\cdots a_1}(\th_n,\dots,\th_1)\,,\sspace (\mbox{$n$ odd})
\label{o3top1}
\ee
where $\cL(p) = - p^2 + m^2$. The subscript `$0$' indicates that under an 
O$(3)$ rotation these functions are interwtiners ${\bf 3}^{\otimes n}\ra
{\bf 1}$ onto the singlet (isospin zero) irreducible representation. 
The normalization of the topological charge operator between physical
states of fixed particle number is not known a priori. In section 5 
we shall determine the non-perturbative normalization constant to be 
\be
\lambda_0=\frac{1}{4}\,.
\label{lambda0}
\ee

Introducing the squares
\be
F_0^{(n)}(u)=\sum_{a_1\dots a_n}
\left|f_{0,a_n\cdots a_1}(\th_n,\dots,\th_1)\right|^2
\label{o3top2}
\ee
we can write the spectral density for this operator as
\begin{subeqnarray}
\rho^{\rm top}(\mu) \is \lambda_0
\sum_{k=0}^\infty\rho^{(2k+1)}_0(\mu)\,,\\
\rho^{(n)}_0(\mu) \is \int_0^{\infty}
\frac{d u_1 \ldots d u_{n-1}}{(4 \pi)^{n-1}}\;F_0^{(n)}(u)\; 
\delta(\mu - M^{(n)}(u))\,.  
\label{o3top3}
\end{subeqnarray}
Finally if we define $I^{\rm top}$ similarly as in (\ref{o36}), the Euclidean
two-point function of the Euclidean topological charge density is given as
\be
\bra k(x)\,k(y) \ket
=-\, \int\frac{d^2 p}{(2\pi)^2}\, e^{-ip\cdot(x-y)}\,
(p^2+m^2)^2\,
I^{\rm top}(p^2)\,.
\label{o3top5}
\ee
Perturbation theory predicts
\ba
p^2\, I^{\rm top}(p^2) \is \iota^{\rm top}_0\, -\,\frac{\alpha(p)}{16\pi}+
O\left(\alpha^2(p)\right)\,,\cr
\mu\,\rho^{\rm top}(\mu) \is \frac{1}{16\pi}\left\{\alpha^2(\mu)+
O\left(\alpha^3(\mu)\right)\right\}\,,
\label{o3top6}
\ea
where $\iota^{\rm top}_0$, being the coefficient of a contact term in the 
two-point function, cannot be calculated in PT. 

{\em Energy-momentum tensor:} Since we already discussed the EM tensor 
in section 2.1, here we can be brief. From the Lagrangian one obtains
\be
T_{\mu\nu}=\frac{1}{g_0^2}\left\{ \d_\mu S^a \d_\nu S^a -
\frac{1}{2}\,\eta_{\mu\nu}\,
\d_\alpha S^a \d^\alpha S^a\right\}\,.
\label{o3EM1}
\ee
In the QFT we fix the normalization such that the Hamiltonian
$\int_{-\infty}^{\infty} dx T_{00}(x)$ has eigenvalue $\sqrt{p^2 +m^2}$
on an asymptotic single particle state of momentum $p$. The scalarized 
form factors can be identified with that of the EM-potential $\tau$ as 
introduced in (\ref{EM1}). To fix the notation we repeat their defining 
relation 
\be
\bra 0|T_{\mu\nu}(0)|a_n,\th_n;\dots;a_1,\th_1\ket = 
\cL_{\mu\nu}(P^{(n)}(\th))\;
f_{0,a_n\cdots a_1}(\th_n,\dots,\th_1)\,,\sspace (\mbox{$n$ even})
\label{o3EM2}
\ee
with $\cL_{\mu\nu}(p)$ as in (\ref{EM1}). Again we can use the same symbol 
here as for the scalarized form factors of the topological charge density 
since the latter are non-vanishing only for odd particle numbers. 
The above normalization of $T_{\mu\nu}$ corresponds to the following 
constraint on the scalarized two-particle form factor
\be
f_{0,a_2a_1}(\th_2,\th_1) =
\frac{-2\delta^{a_2a_1}}{(\th_2-\th_1-i\pi)^2} +\ldots \;,
\;\;\; \th_2\approx\th_1+i\pi\;.
\label{o3EM4}
\ee
If we now extend the definitions (\ref{o3top2}) and (\ref{o3top3}b) for
even values of $n$, we can write
\be
\rho^{\rm EM}(\mu) = \sum_{k=1}^\infty\rho^{(2k)}_0(\mu)\,,
\label{o3EM6}
\ee
and define $I^{\rm EM}$ in terms of $\rho^{\rm EM}(\mu)$ as in 
(\ref{sp10}). The various two-point functions can then be computed 
along the lines described in section 2.1.

Finally the perturbative results for the EM tensor are:
\ba
p^2\, I^{\rm EM}(p^2) \is \frac{1}{6\pi} -\,\frac{\alpha(p)}{4\pi}+
O\left(\alpha^2(p)\right)\,,\cr
\mu\,\rho^{\rm EM}(\mu) \is \frac{1}{4\pi}\left\{\alpha^2(\mu)+
O\left(\alpha^3(\mu)\right)\right\}\,.
\label{o3EM7}
\ea
In addition, the integral of the spectral density is constrained by 
(\ref{AS3}):
\be
\int_0^\infty\, d\mu\,\rho^{\rm EM}(\mu)=\frac{1}{6\pi}\,,
\label{o3EM8}
\ee
expressing the fact that the UV central charge of the O$(3)$ model 
is equal to 2.

%%%%%%%%%%%%%%%%%%%%%%%%%%%%%%%%%%%%%%%%%%%%%%%%%%%%%%%%%%%%%%%%%%%%%%%
%\include{sect3}
\newsection{Results for the O$(3)$ form factors}

In this section we collect our results for various O$(3)$ form factors,
form factor squares and their properties
and we give an exact expression 
for the asymptotics of the $n$-particle spectral densities. 
We refrain from going into the details of the derivation and giving proofs,
which can be found in \cite{BN2}.

The most remarkable feature of the NLS model is its integrability.
Assuming asymptotic freedom one can establish the existence of  
non-local conserved charges and the absence of particle production
\cite{Luscher}. The matrix elements of the non-local charges between  
physical particle states together with the O$(3)$ symmetry determine 
the S-matrix up to a CDD ambiguity, and the result agrees with that 
obtained from the S-matrix bootstrap \cite{Zam}. For an overview of these
issues see also \cite{AAR}. The O$(3)$ bootstrap S-matrix 
has been tested (at low energies) in lattice studies \cite{LW} and used 
as an input for the thermodynamic Bethe Ansatz (TBA) to compute the exact 
$m/\Lambda$ ratio \cite{HN}. As a by-product of the TBA considerations
the CDD ambiguity can be resolved. 
The construction of the non-local charges has been extended to
the full quantum monodromy matrix in \cite{DeVega}.
More recently the algebraic structure underlying these 
non-local charges has been considerably generalized \cite{BeLeCl} and 
has also been identified directly in the context of the form factor 
bootstrap \cite{MN}. 

The Form Factor Bootstrap method was initiated in \cite{Weiszetal} and 
has been further developed by Smirnov \cite{KirSmir,Smir}. Using the exact
S-matrix as an input the form factors get constructed as solutions of a 
system of functional equations known as Smirnov axioms. These equations 
recursively relate the $n$-particle form factors to the $(n-2)$-particle 
form factors and (provided the S-matrix also has bound state poles) to the
$(n-1)$-particle form factors. The method is reviewed in Smirnov's book
\cite{Smir} and, for the special case of the O(3) model, in
\cite{B,BH}.

\newsubsection{Reduced form factors}

Let us recall the
simplest nontrivial case, i.e. the result for the
2-particle scalarized form factor of the Noether current \cite{Weiszetal}:
\be
f^a_{1,a_2a_1}(\th_2,\th_1)=
\frac{\pi^2}{2}\,\psi(\th_2-\th_1)\,\epsilon^{aa_2a_1}\,,
\label{FF1}
\ee
where
\be
\psi(\th)=\frac{\th-i\pi}{\th(2\pi i-\th)}\tanh^2\frac{\th}{2}\,.
\label{FF2}
\ee
\vspace{-1mm}

Here the normalization condition (\ref{o319}) has been taken into account.
The product of the 2-particle solutions for all possible pairs of rapidity
differences,
\be
\Psi(\th_n,\dots,\th_1)=\prod_{i>j}\psi(\th_i-\th_j)\,,
\label{FF3}
\ee
will play an important role in what follows. Actually, if there were
no internal indices, (\ref{FF3}) would be a solution for the 
$n$-particle case. All the following complications (absent in integrable 
models with diagonal S-matrices) are due to the presence of 
these internal quantum numbers. Using (\ref{FF3}) we can define the reduced
form factors $g^a_{1,a_n\cdots a_1}(\th_n,\dots,\th_1)$ for the
Spin \& Current operators as
\be
f^a_{1,a_n\cdots a_1}(\th_n,\dots,\th_1)=\frac{\pi^{\frac{3n}{2}-1}}{2}
\Psi(\th_n,\dots,\th_1)\,
g^a_{1,a_n\cdots a_1}(\th_n,\dots,\th_1)
\label{FF4}
\ee
and similarly for the TC density \& EM tensor operators as
\be
f_{0,a_n\cdots a_1}(\th_n,\dots,\th_1)=\frac{\pi^{\frac{3n}{2}-1}}{2}
\Psi(\th_n,\dots,\th_1)\,
g_{0,a_n\cdots a_1}(\th_n,\dots,\th_1)\,.
\label{FF5}
\ee
\vspace{1mm}

The first few reduced form factors are
\begin{subeqnarray}
&&g^a_{1,a_1}(\th_1)=\delta^{a_2a_1}\,,\\
&&g^a_{1,a_2a_1}(\th_2,\th_1)=\epsilon^{aa_2a_1}\,,\\
&&g_{0,a_2a_1}(\th_2,\th_1)=\frac{\delta^{a_2a_1}}{\th_2-\th_1-i\pi}\,,\\
&&g_{0,a_3a_2a_1}(\th_3,\th_2,\th_1)=\epsilon^{a_3a_2a_1}\,,
\label{FF6}
\end{subeqnarray}
where the normalization conditions (\ref{o31}), (\ref{o319}) and (\ref{o3EM4}) 
have been used. Note that we have not found the normalization 
condition for the topological charge density operator yet. 
Its lowest reduced form factor (\ref{FF6}d) was chosen here to satisfy 
the clustering relations in the form described below. The still unknown 
normalization of the physical operator is given by (\ref{o3top1}).

What makes the reduced form factors useful is the fact that
(with the notable exception of the 2-particle form factor of the EM
tensor (\ref{FF6}c)) they are all polynomial in the rapidities,
with specified total degree $N(n)$ and partial degree $p(n)$ given by
\cite{BHN}
\be
N(n) = \frac{1}{2}(n^2 -3n) +l\;,\sspace p(n) = n-3 + l\;,\sspace l =0,1\;.
\label{FF7}
\ee
By partial degree we mean the degree in an individual rapidity variable
and $l=0,1$ correspond to the isospin 0 and 1 families, respectively.
Expressions for the O$(3)$ form factors were first written down in 
\cite{KirSmir}. Exploiting their polynomiality and the intertwining 
property, they were recast in \cite{B,BH} into a form more suitable 
for practical purposes. The fact that suitable reduced form factors are 
polynomials in the rapidities is a special feature of the O$(3)$ model 
that fails to hold for the higher O$(N),\;N>3$ models and which
considerably facilitates their explicit construction. We have computed the 
form factors for both 
families of operators (Spin \& Current and TC density \& EM tensor) up to 
6 particles. The complexity of these reduced form factors grows very
rapidly with the particle number, because the number of components 
(with respect to the internal quantum numbers), the number of independent 
rapidity differences and the degree of the polynomials all grow very fast 
with increasing particle number. For 5- and 6-particles the polynomials 
are already too large to be communicated in print.%
\footnote{For example the $6$-particle polynomial of the current has
about 3.4 Mbytes, i.e. about 700 A4 pages.} 
Below we complete the list (\ref{FF6}) for up to 4 particles:

Spin \& Current:
\bas
&n=3:& g^a_{1,a_3a_2a_1}(\th) = 
\delta^{aa_1}\delta^{a_3a_2}\,g_1(\th)+
\delta^{aa_2}\delta^{a_3a_1}\,g_2(\th)+
\delta^{aa_3}\delta^{a_2a_1}\,g_3(\th)\;,\nonum
&&
\left( \begin{array}{c} 
g_1(\th) \\ g_2(\th) \\ g_3(\th) \end{array} \right)\! =
\left( \begin{array}{c} 
\th_{23} \\ \th_{31} -2 i\pi \\ \th_{12}\end{array} \right)\;.\nonum
&n=4:& g^a_{1,a_4a_3a_2a_1}(\th) = 
\delta^{a_4a_3}\epsilon^{aa_2a_1}\,g_1(\th) +
\delta^{a_4a_2}\epsilon^{aa_3a_1}\,g_2(\th) +
\delta^{a_4a_1}\epsilon^{aa_3a_2}\,g_3(\th) \nonum
&&\sspace\bspace\!\!\!\! +\delta^{a_3a_2}\epsilon^{aa_4a_1}\,g_4(\th) +
\delta^{a_3a_1}\epsilon^{aa_4a_2}\,g_5(\th) +
\delta^{a_2a_1}\epsilon^{aa_4a_3}\,g_6(\th)\;,\nonum 
&& 
\left( \begin{array}{c} 
g_1(\th) \\ g_2(\th) \\ g_3(\th) \\ g_4(\th) \\ 
g_5(\th) \\ g_6(\th)
\end{array} \right)\! =
\left( \begin{array}{c} 
-i\pi(u^2 + v^2 -i\pi u - i\pi v + 2\pi^2) \\
(u-i\pi)v(v-i\pi)\\
(u-i\pi)(u+2i\pi)(i\pi -v) \\ 
uv(3i\pi - v)\\ 
u(u-i\pi)v \\ 
2i\pi(i\pi-u)v 
\end{array} \right) \nonum
&& 
+ (\th_{43} -i\pi) \left( \begin{array}{c}
-4\pi^2 -i\pi(u+v) -(u-v)^2\\ 
-2\pi^2 - 3i\pi v +v^2\\
-4\pi^2 + i\pi(u - 2 v) - u^2 \\ 
2 \pi^2 + i\pi(u + 2 v) - 2 u v \\
-i\pi(2 u + v) + 2uv\\ 
-2\pi^2 + i\pi(u - 3 v) 
\end{array} \right)+ 
(\th_{43} -i\pi)^2 \left( \begin{array}{c}
0\\
0\\
0\\
-u\\
v-2i\pi\\
u-v
\end{array} \right)\;,\nonum
&& \mbox{where $u = \th_{32},\;v=\th_{31}$.}
\eas
EM tensor \& TC density:
\bas
&n=4:& g_{0,a_4a_3a_2a_1}(\th) = 
g_1(\th) \delta^{a_1a_2}\delta^{a_3a_4} +
g_2(\th) \delta^{a_4a_2}\delta^{a_3a_1} +
g_3(\th) \delta^{a_4a_1}\delta^{a_3a_2} \;,\nonum
&& 
\left( \begin{array}{c}
 g_1(\th) \\ g_2(\th) \\ g_3(\th)
\end{array} \right) = 
\left( \begin{array}{c}
-\th_{21}\th_{43}+2 \pi^2 \\
\th_{32}\th_{41}+\th_{21}\th_{43}-2i\pi(\th_{41}-i\pi)\\
-\th_{32}\th_{41}+2i\pi(\th_{32}+i\pi)
\end{array} \right)\;.
\eas
Here we used the shorthand notation $\th_{ij}$ for $\th_i-\th_j$.

%%%%%%%%%%%%%%%%%%%%%%%%%%%%%%%%%%%%%%%%%%%%%%%%%%%%%%%%%%%%%%%%%%%%

\newsubsection{Clustering properties}

A remarkable property of the form factors is asymptotic clustering. If we
divide the set of rapidities into two disconnected clusters and boost
one of them with respect to the other, for asymptotically large boosts the
form factors factorize into a sum of products of form factors corresponding
to the two clusters separately. In formulae
\ba
&&g^a_{1,a_k\cdots a_1b_m\cdots b_1}(\alpha_k+\Delta,\dots,\alpha_1+\Delta,
\beta_m,\dots,\beta_1)\nonum
&&\sspace =\Delta^{km-1}\epsilon^{abc}
g^b_{1,a_k\cdots a_1}(\alpha_k,\dots,\alpha_1)\, g^c_{1,b_m\cdots b_1}(\beta_m,
\dots,\beta_1)+O(\Delta^{km-2})
\label{CL1}
\ea
and similarly
\ba
&&g_{0,a_k\cdots a_1b_m\cdots b_1}(\alpha_k+\Delta,\dots,\alpha_1+\Delta,
\beta_m,\dots,\beta_1)\nonum
&&\sspace=\Delta^{km-2}
g^a_{1,a_k\cdots a_1}(\alpha_k,\dots,\alpha_1)\, g^a_{1,b_m\cdots b_1}(\beta_m,
\dots,\beta_1)+O(\Delta^{km-3})\,.
\label{CL2}
\ea
Note that in (\ref{CL1}) members of the isospin 1 family are mapped onto
themselves, while in (\ref{CL2}) members of the isospin 1 family are
linked to members of the isospin 0 family. Observe also that there is
no distinction between even and odd members of the same family, their
factorization properties are the same. Thus even and odd members of the two
families are very closely related. We already anticipated this close
interrelation (again a special O$(3)$ property) by using the same notation
for even and odd form factors and we introduced the non-perturbative
constants $2/\sqrt{\pi}$ and $\lambda_0$ in (\ref{o33}) and (\ref{o3top1}) 
respectively in order to have this simple form for (\ref{CL1}) and 
(\ref{CL2}). 

For the special case of $k=1$, $m=n-1$ the clustering relations read
\ba
&& g^a_{1,a_n\cdots a_1}(\th_n,\th_{n-1},\dots,\th_1)
=\th_n^{n-2}\epsilon^{aa_nb}
g^b_{1,a_{n-1}\cdots a_1}(\th_{n-1},\dots,\th_1)+O(\th_n^{n-3})\,, \nonum
&& g_{0,a_n\cdots a_1}(\th_n,\th_{n-1},\dots,\th_1)
=\th_n^{n-3}
g^{a_n}_{1,a_{n-1}\cdots a_1}(\th_{n-1},\dots,\th_1)+O(\th_n^{n-4})\,,
\label{CL3}
\ea
from which one sees that the reduced form factors are polynomials of partial 
degree $(n-2)$ and $(n-3)$ in the isospin 1 and 0 case, respectively.

Since the product (\ref{FF3}) also factorizes under clustering, the full
(scalarized) form factors also satisfy clustering relations, which are similar
to (\ref{CL1}) and (\ref{CL2}). For the $l=1$ family they
can be found in Smirnov's book \cite{Smir}. Similar clustering
relations were recently discussed in \cite{CL}.

The clustering relations closely resemble some classical equations
satisfied by our operators. For example dividing an even number of 
particles into two odd clusters, (\ref{CL1}) can be interpreted as 
the quantum counterpart of (\ref{o315}),
the classical definition of the current in terms of the spin operators.
(Remember that we are dealing with scalarized objects so that all 
information on the Lorentz structure is lost.) The division
of an even number of particles into two even clusters, on the other hand,
resembles the classical relation
\be
\d_\mu J^a_\nu -\d_\mu J^a_\mu=2g_0^2\epsilon ^{abc}J^b_\mu J^c_\nu\,.
\label{CL5}
\ee
Finally the clustering of an odd number of particles
corresponds to
\be
\d_\mu S^a=-g_0^2\epsilon^{abc}S^b J^c_\mu\,.
\label{CL6}
\ee
Similarly, (\ref{CL2}) corresponds to the defining equation (\ref{o3EM1}) or
either of
\be
T_{\mu\nu}=g_0^2\{J^a_\mu J^a_\nu-\frac{1}{2}\eta_{\mu\nu}J^a_\alpha
J^{a\alpha}\}\;,\sspace 
q=\frac{g_0^2}{8\pi}\epsilon^{\mu\nu} J^a_\mu \d_\nu S^a\,.
\label{CL7}
\ee

%%%%%%%%%%%%%%%%%%%%%%%%%%%%%%%%%%%%%%%%%%%%%%%%%%%%%%%%%%%%%%%%%%%%%%%%
\newsubsection{Form factor squares}

The quantity entering the spectral densities and two-point functions is   
the absolute square of the form factors, summed over the internal symmetry
indices. For the reduced form factors the corresponding quantities
are
\ba
G_1^{(n)}(\th_n,\dots,\th_1)
\is \frac{1}{3}\sum_{aa_n\cdots a_1}
|g^a_{1,a_n\cdots a_1}(\th_n,\dots,\th_1)|^2\;,\nonum
G_0^{(n)}(\th_n,\dots,\th_1) \is \frac{1}{3}\sum_{a_n\cdots a_1}
|g_{0,a_n\cdots a_1}(\th_n,\dots,\th_1)|^2\,.
\label{SQ1}
\ea
Here are the results for the first few reduced form factor squares:
\ba
G^{(1)}_1(\th_1)&=&1\,,\nonum
G^{(2)}_1(\th_2,\th_1)&=&2\,,\nonum
G^{(3)}_1(\th_3,\th_2,\th_1)&=&2(\th_{32}^2+\th_{31}^2+\th_{21}^2)
+12\pi^2\,,\nonum
G^{(2)}_0(\th_2,\th_1)&=&\frac{1}{\th_{21}^2+\pi^2}\,,\nonum
G^{(3)}_0(\th_3,\th_2,\th_1)&=&2\,,\nonum
G^{(4)}_0(\th_4,\th_3,\th_2,\th_1)&=&2(\th_{43}^2\th_{21}^2+
\th_{42}^2\th_{31}^2+\th_{41}^2\th_{32}^2)\nonum
&&+4\pi^2(\th_{43}^2+
\th_{42}^2+\th_{41}^2+\th_{32}^2+\th_{31}^2+\th_{21}^2)+28\pi^4\,.
\label{SQ2}
\ea

The reduced form factor squares $G^{(n)}_l$ ($l=0,1$) are symmetric, boost
invariant polynomials of total degree $n^2-3n+2l$ and partial degree
$2(n-3+l)$. They satisfy the clustering relations
\ba
&& G_l^{(k+m)}(\alpha_k +\Delta,\ldots,\alpha_1 +\Delta, 
\beta_m,\ldots,\beta_1)\sspace \nonum
&&\ \ \  = (l+1)\,\Delta^{2km-4+2l}\,G_1^{(k)}(\alpha_k,\ldots,\alpha_1)\,
G^{(m)}_1(\beta_m,\ldots,\beta_1) + O(\Delta^{2km-5+2l})\;,
\label{cl1}
\ea
which follow from (\ref{CL1}) and (\ref{CL2}).
Again notice that the members of the Spin \& Current series ($l=1$) are
mapped onto themselves under clustering, while the members of  
the EM tensor \& TC density series ($l=0$) are linked to the $l=1$ series. 

(\ref{cl1}) is readily translated into the corresponding 
statement about the full form factor squares (\ref{o34}) and (\ref{o3top2})
\ba
&& F_l^{(k+m)}(\alpha_k +\Delta,\ldots,\alpha_1 +\Delta, 
\beta_m,\ldots,\beta_1) \sspace \nonum
&&\sspace = \frac{(12-4l)\pi^2}{\Delta^{4-2l}}\,
F_1^{(k)}(\alpha_k,\ldots,\alpha_1)\,
F^{(m)}_1(\beta_m,\ldots,\beta_1) + O(\Delta^{-5+2l})\;.
\label{cl2}
\ea
 
Next we discuss two further properties of the polynomials
$G^{(n)}_l$. First we present an explicit formula for the overall leading
terms of these polynomials, i.e. the terms with the maximal total degree,
which is $n^2-3n+2l$. These overall leading terms are given by
\ba
G^{(n)}_{0;0}(\th_n,\dots,\th_1) \is \frac{1}{n}\,
\left(\prod_{i>j}\th_{ij}^2\right)\,\sum_{\sigma\in S_n}
P_0(\th_{\sigma(n)},\dots,\th_{\sigma(1)})\,,\nonum
G^{(n)}_{1;0}(\th_n,\dots,\th_1) \is
\left(\prod_{i>j}\th_{ij}^2\right)\,\sum_{\sigma\in S_n}
P_1(\th_{\sigma(n)},\dots,\th_{\sigma(1)})\,.
\label{overall}
\ea
Here the summations range over all elements of the permutation group 
$S_n$ and the functions $P_0$ and $P_1$ are defined as
\be
P_0(\th_n,\dots,\th_1)=\frac{1}{\th_{n\, n-1}^2\cdots\th_{32}^2\th_{21}^2
\th_{1n}^2}\;,\sspace 
P_1(\th_n,\dots,\th_1)=\frac{1}{\th_{n\, n-1}^2\cdots\th_{32}^2\th_{21}^2}\;,
\label{P01}
\ee
respectively. These overall leading terms $G^{(n)}_{l;0}$ by 
themselves satisfy the clustering relations (\ref{cl1}).  
An equivalent expression for the leading terms (\ref{overall}) was 
obtained by H.~Lehmann \cite{Leh}.  

The second property concerns the analytic continuation of the squares 
$G^{(n)}_l$ to complex rapidities. For real, physical rapidities the 
polynomials $G^{(n)}_l$  are real-valued and positive. Being polynomials 
one can also evaluate them for complex rapidities $\th_i$. In 
particular they turn out to vanish if two consecutive rapidity differences 
are both equal to 
$i\pi$
\be
G^{(n)}_l(\th_{n-2}+2\pi i,\th_{n-2}+i\pi,\th_{n-2},\dots,\th_1)=0\,,
\sspace n\geq 4-l\,.
\label{Rsub}
\ee
$G^{(n)}_l$ vanishes also at all other points in rapidity space 
obtained from the one in (\ref{Rsub}) by permutation of the arguments.

Since the form factor squares -- not the form factors themselves -- are 
the objects entering the physically interesting quantities through the 
spectral representation it would be desirable to obtain them directly, 
without having to go through the tedious procedure of first computing 
the form factors and then squaring them. Indeed, for the first few $n$ 
the information collected in this subsection is sufficient to determine 
$G^{(n)}_l$ directly. However, for particle numbers 5 and 6, although these 
constraints -- permutation symmetry, boost invariance, 
given total and partial degrees, given overall leading terms, clustering 
and the vanishing at points (\ref{Rsub})  -- almost uniquely determine 
the solution, but not quite. So in these cases we had to work out the 
solution the tedious way. We hope to find additional general properties of 
the squares that will determine them completely. In the appendix we 
list the results for the form factor squares for the EM tensor \& 
TC density and Current \& Spin series up to 6 particles.

%%%%%%%%%%%%%%%%%%%%%%%%%%%%%%%%%%%%%%%%%%%%%%%%%%%%%%%%%%%%%%%%%%%%%%%%%
\newsubsection{Asymptotics of spectral densities}

Let us now turn to the computation of the spectral densities 
for the four operators under 
consideration. The $n$-particle contributions are
\ba
\rho_l^{(n)}(\mu) \is \int_0^{\infty}
\frac{du_1\ldots du_{n-1}}{(4\pi)^{n-1}}\,
F_l^{(n)}(u) \;\delta(\mu - M^{(n)}(u))\;,\;\;\;n \geq 2, \nonum
F^{(n)}_l(u) \is\frac{\pi^{3n-2}}{4}\,(3-2l)\, G^{(n)}_l(u)\,|\Psi(u)|^2\,.
\label{asy1}
\ea
The integration in (\ref{asy1}) can be done numerically. 
The graph of the resulting $n$-particle spectral densities is roughly 
`bell-shaped': 
Starting from zero at $\mu = m n$  they are strictly increasing,  
reach a single maximum and then decrease monotonically for  
large $\mu$. Some typical plots can be found in \cite{BHN}.

Of particular interest is the $\mu \ra \infty$ asymptotics of the
spectral densities. One can show that an important consequence of the 
clustering relations (\ref{cl2}) is the following expression for
the asymptotic form of the $n$-particle spectral densities
\be 
\rho^{(n)}_l(\mu) \,\sim\, \frac{A^{(n)}_l}{\mu (\ln\mu)^{4-2l}}\;,\sspace
\mu \ra \infty\;,
\label{asy2}
\ee
where the constants $A^{(n)}_l$ are computable from the integrals 
of the {\em lower} particle spectral densities. The integrals are
\be
c^{(n)}_1=\int_0^\infty\, d\mu\,\rho^{(n)}_1(\mu)=\frac{1}{\pi}
\left(\frac{\pi}{2}\right)^{2n}\int_0^\infty\, du_1\dots du_{n-1}\,
G^{(n)}_1(u)\,|\Psi(u)|^2 \;,\;\;n\geq 2\,, 
\label{asy3}
\ee
and in terms of them the decay constants $A^{(n)}_l$ are given by
\be
A^{(n)}_1=\frac{8}{3}\,A^{(n)}_0=\pi\sum_{k=1}^{n-1}\,c^{(k)}_1\,c^{(n-k)}_1
\,, \sspace n\geq2\,,
\label{asy4}
\ee
if (\ref{asy3}) is supplemented by the definition $c^{(1)}_1:=\pi/4$.
Note that (\ref{asy4}) implies that the $A_l^{(n)}$ are strictly 
increasing with $n$. This implies that for sufficiently high energies
the $n$-particle contribution overtakes the $(n-2)$-particle 
contribution, i.e. $\rho^{(n)}(\mu)> \rho^{(n-2)}(\mu)$, for 
$\mu > \mu_n$, where $\rho^{(n)}(\mu_n)= \rho^{(n-2)}(\mu_n)$;
although the maximum of $\rho^{(n)}(\mu)$ is expected to be  
much smaller than the maximum of $\rho^{(n-2)}(\mu)$ (c.f.~section 4). 
This is the ``crossover phenomenon'' observed in \cite{BHN} for low 
particle numbers. 
Although (\ref{asy2}) is an exact asymptotic equation, the way how the 
asymptotic behavior is approached is highly non-uniform in the particle 
number. That is to say, with increasing $n$ one has to go to larger and 
larger $\mu$ in order to make the
right hand side of (\ref{asy2}) a good description of the function  
$\rho^{(n)}_l(\mu)$. 

A consequence of the crossover is that the asymptotic behavior of the 
exact spectral density (being the sum of all even/odd $n$-particle 
contributions) {\em cannot} be obtained by naively summing up the right 
hand sides of (\ref{asy2}), which in fact would be divergent. This 
divergence signals that the infinite series has a more singular high 
energy behavior than each of its terms separately. 
Indeed, according to PT,
the high energy behavior of the full spectral densities is  
\ba
\cO = \mbox{EM \& top:}&\sspace & \;\; 
\rho(\mu) \,\sim\, \frac{A^{\cO}}{\mu}\left[\frac{1}{(\ln\mu)^2} +
O\left(\frac{\ln\ln\mu}{(\ln\mu)^3}\right)\right]\,,\nonum
\cO = \mbox{spin \& curr:}&\sspace& \;\;
\rho(\mu) \,\sim\, \frac{A^{\cO}}{\mu}\left[1+ O\left(
\frac{1}{\ln \mu}\right)\right]\,,
\label{asy5}
\ea
where the values of the various decay constants can be read off from
the equations (\ref{o313}), (\ref{o323}), (\ref{o3top6}) and
(\ref{o3EM7}). Clearly a finite number of terms, each decaying according to
(\ref{asy2}) can never produce the more singular UV behavior of (\ref{asy5}).
It is therefore a major challenge to develop re-summation techniques for
the spectral resolution and, for the reasons outlined in the introduction, 
compute the UV properties independent of PT. We shall meet this challenge 
in section 5.

%%%%%%%%%%%%%%%%%%%%%%%%%%%%%%%%%%%%%%%%%%%%%%%%%%%%%%%%%%%%%%%%%%%%%%%%%
%\include{sect4}
%\newpage
\newsection{Results for the two-point functions}

Knowing the form factor squares the evaluation of the $n$-particle 
contributions to the spectral densities and the two-point functions is 
in principle straightforward. For the spectral densities the 
integrations in (\ref{o35}b) and (\ref{o3top3}b) can be done numerically 
to good accuracy. Throughout we used an accuracy of $10^{-3}$ for all
numerical computations. 
For comparison with PT and MC data it is useful to consider the Fourier 
transform $I(p)$ of the two-point function, which can be computed from
the spectral density via (\ref{sp10}). Of course $I(p)$ again decomposes
into a sum of $n$-particle contributions of which in practice only the 
first few are known explicitly. It is important that an {\em intrinsic} 
estimate for the error induced by this truncation can be made, i.e.~one 
which does not rely on comparison with other techniques. 

To discuss this qualitative error estimate let us first note some general 
features of the $n$-particle contributions to the spectral densities.
As remarked before the graph of an $n$-particle spectral
density is roughly bell-shaped: Starting from zero at $\mu = mn$ it is 
strictly increasing, reaches a single maximum and then decreases 
monotonically, the $\mu \ra \infty$ asymptotics being given by
(\ref{asy2}). With increasing particle number $n$ the values of the maxima 
rapidly decrease. Generally speaking the maximum of $\rho^{(n)}(\mu)$ is 
smaller by 1.5 to 2.5 orders of magnitudes compared to the maximum of 
$\rho^{(n-2)}(\mu)$, while its position is shifted to higher energies. 
Nevertheless at sufficiently high energies the $n$-particle contribution 
overtakes the $(n-2)$-particle contribution, i.e. $\rho^{(n)}(\mu)> 
\rho^{(n-2)}(\mu)$, iff $\mu > \mu_n$, where $\rho^{(n)}(\mu_n)= 
\rho^{(n-2)}(\mu_n)$ \cite{BHN}. For $n\leq 6$ the positions and the 
values of the maxima are listed in Table 1 below. The results for the
points of intersection  
$\big(\mu_n/m,\,\rho^{(n-2)}(\mu_n) = \rho^{(n)}(\mu_n)\big)$ are as follows:
\ba
\mbox{Spin:}\;\;&\mbox{$n=5$:}\;\;\;(1.0\cdot 10^4, 4.9\cdot 10^{-6})\;,\;\;
&\nonum
\mbox{Current:}\;\;&\mbox{$n=4$:}\;\;\;(1.6\cdot 10^2,3.9\cdot10^{-4})\;,\;\;
&\mbox{$n=6$:}\;\;\;(1.0\cdot 10^6, 3.9\cdot 10^{-8})\;,
\nonum
\mbox{TC:}\;\;&\mbox{$n=5$:}\;\;\;(7.0\cdot 10^3,3.8\cdot10^{-8})\;,\;\;
&\nonum
\mbox{EM:}\;\;&\mbox{$n=4$:}\;\;\;(1.6\cdot 10^2, 4.7\cdot 10^{-6})\;,\;\;
&\mbox{$n=6$:}\;\;\;(4.6\cdot 10^5, 2.5\cdot 10^{-10})\;.
\ea
It is natural to assume that the general trend depicted by these numbers 
continues to hold at higher particle numbers; a quantitative form 
of this assumption will be presented in section 5. Here already the 
qualitative features ($\mu_n$ increasing; maxima decreasing) are 
sufficient to conclude that the point of intersection 
$\mu_n$ provides an intrinsic measure for the quality of the approximation 
made by truncating the form factor series at the $n$-particle term: Since 
$\mu_{n+2} \gg \mu_n$ the $(n+2)$-particle contribution can safely be 
ignored up to energies $\mu \lsim \mu_n$ in the sense that the correction 
to $\rho^{(2)}(\mu) +\ldots + \rho^{(n)}(\mu)$ for $\mu\lsim\mu_n$ should 
be at most a few per mille. 
Thus, the form factor series truncated at 6 particles should provide 
results accurate to within a few per mille for the spectral density at 
least up to energies $7\cdot 10^3 m$ for Spin and TC and $5\cdot 10^5 m$ 
for Current and EM. 
It is not hard to see that for the functions $I(p)$ related to $\rho(\mu)$ 
by the Stieltjes transform (\ref{sp10}) this implies accurate results for 
somewhat smaller ranges, about $p \lsim 10^3$ for the Spin and TC cases
and about $p \lsim 10^4$ for the Current and EM tensor. 
Both of these energy/momentum 
ranges exceed that accessible through MC simulations by several orders 
of magnitudes.

\begin{figure}[htb]
\begin{center}
%\vskip 10mm
\leavevmode
\epsfxsize=170mm
\epsfbox{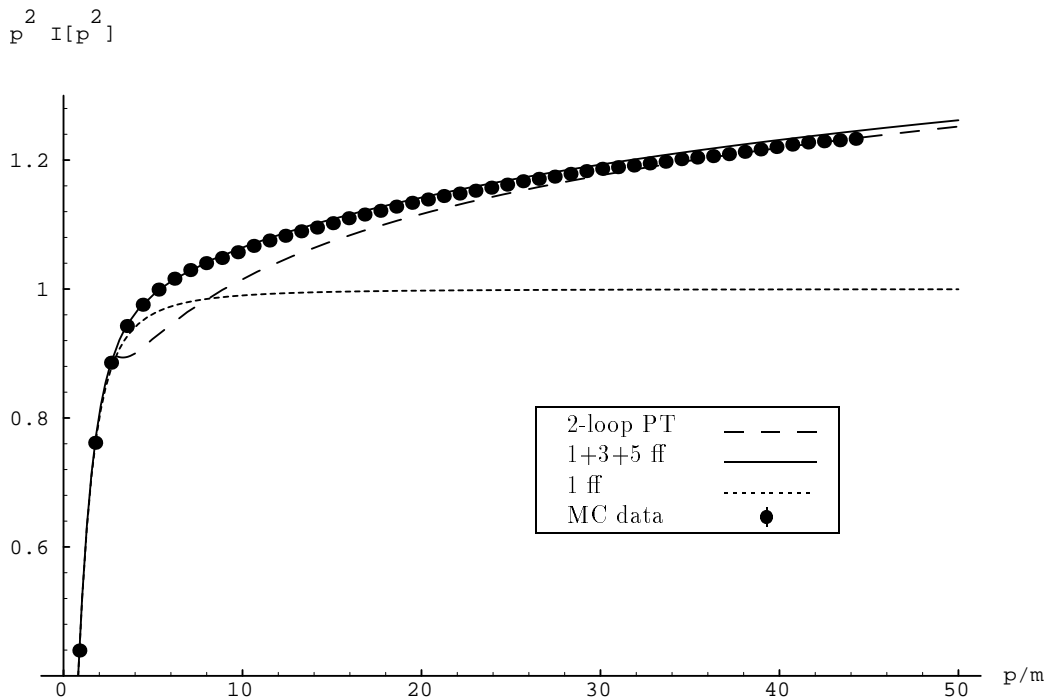}
%\vskip 10mm
\end{center}
\caption{Low energy region of spin two-point function. Comparison: 
Form factor approach, Monte Carlo data and 2-loop perturbation theory; 
$p^2 I(p^2)$ plotted against $p/m$. The normalization of the PT curve is
fixed according to (\ref{lb1}).} 

\label{spin1}
\end{figure}

\begin{figure}[htb]
\begin{center}
%\vskip 10mm
\leavevmode
\epsfxsize=170mm
\epsfbox{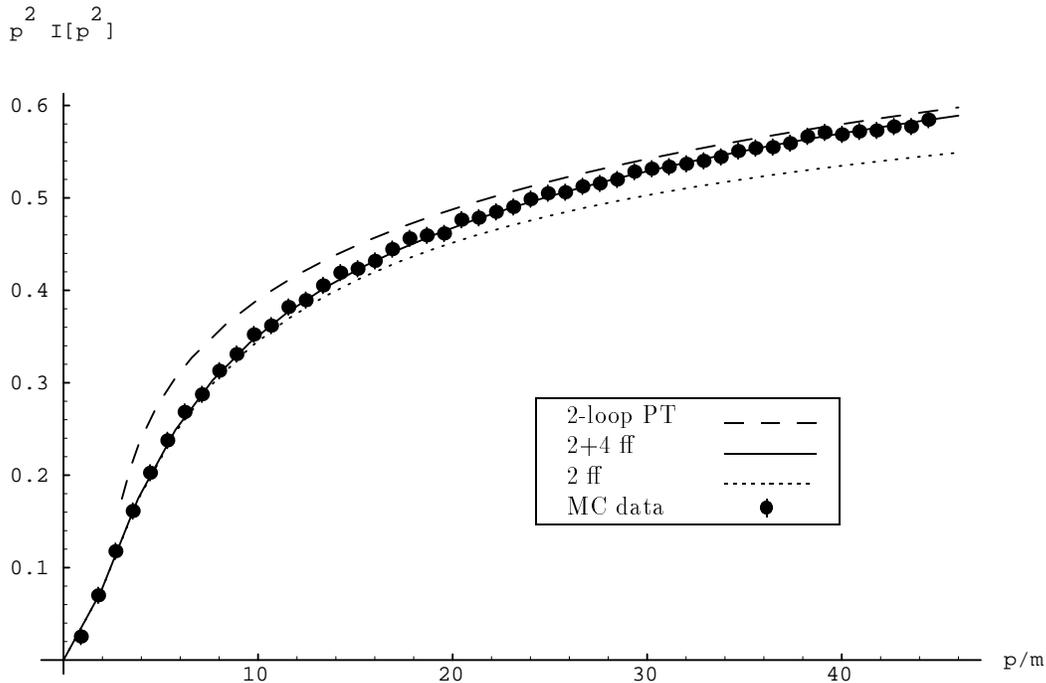}
%\vskip 10mm
\end{center}
\caption{Low energy region of current two-point function. Comparison: 
Form factor approach, Monte Carlo data and 2-loop perturbation theory; 
$p^2 I(p^2)$ plotted against $p/m$.} 
\label{curr1}
\end{figure}

Below we shall restrict attention to the Spin field and 
the Noether current. In both cases it is convenient to consider 
the low- and the high energy region separately. In the low energy 
region non-perturbative effects are expected to become important. Here
Monte Carlo simulations provide an alternative non-perturbative technique 
to probe the system \cite{LW,LWW,MC,PSMC}. Simulations for the two-point 
functions were made \cite{PSMC} using a Wolff-type cluster algorithm 
\cite{UWolff} on a 460 square lattice at inverse 
coupling $\beta = 1.80$ (correlation length $\xi = 65.05$). 
A comparison between form factor results, MC data and PT in this 
low energy region is shown in Figures \ref{spin1} and \ref{curr1} 
for the Spin and Current case, respectively. 

%%%%%%%%%%%%%%%%%%%%%%%%%%%%%%%%%%%%%%%
% figure spin1
%%%%%%%%%%%%%%%%%%%%%%%%%%%%%%%%%%%%%%%%%

%%%%%%%%%%%%%%%%%%%%%%%%%%%%%%%%%%%%%%%
% figure curr1
%%%%%%%%%%%%%%%%%%%%%%%%%%%%%%%%%%%%%%%%%

The agreement between the MC data and the form factor results is
excellent. In the Spin case the statistical errors of the MC data are 
less than the size of the dots in Figure \ref{spin1}. Above $30m$ they
have a slight tendency to lie below the form factor curve, which
(comparing e.g.~with the data at $\beta =1.7$) can be attributed to the still
finite correlation length. The $1+3$ ff curve is left out in Figure 1 as 
it coincides with the $1+3+5$ ff curve below $30m$ and (incidentally) with 
the PT curve above 40m. In the Current case the statistical errors 
are larger but within the errors the agreement with the form factor curve
is perfect. One also sees that for energies between 30 and 45 the PT 
curve runs almost parallel to the MC data in Figure \ref{curr1}. 
Without the guidance of the form factor result one would 
thus be tempted to match both curves by tuning the Lambda parameter 
appropriately. Doing this however, the Lambda parameter comes out wrong by 
about 10\% (from below), as compared with the known exact result \cite{HN}. 
Generally speaking one sees that a determination of the Lambda parameter
(accurate to within 1\% say) from MC data and (2-loop) PT alone is 
difficult because some assumption about the onset of the 2-loop
perturbative regime enters. It remains difficult even after
3-loop effects are taken into account \cite{Shin}. To investigate the onset
of a perturbative regime let us now consider the high energy regime of the 
same quantities.  

%%%%%%%%%%%%%%%%%%%%%%%%%%%%%%%%%%%%%%%
% figure spin2
\begin{figure}[htb]
\begin{flushleft}
%\vskip 10mm
\leavevmode
\epsfxsize=170mm
\epsfbox{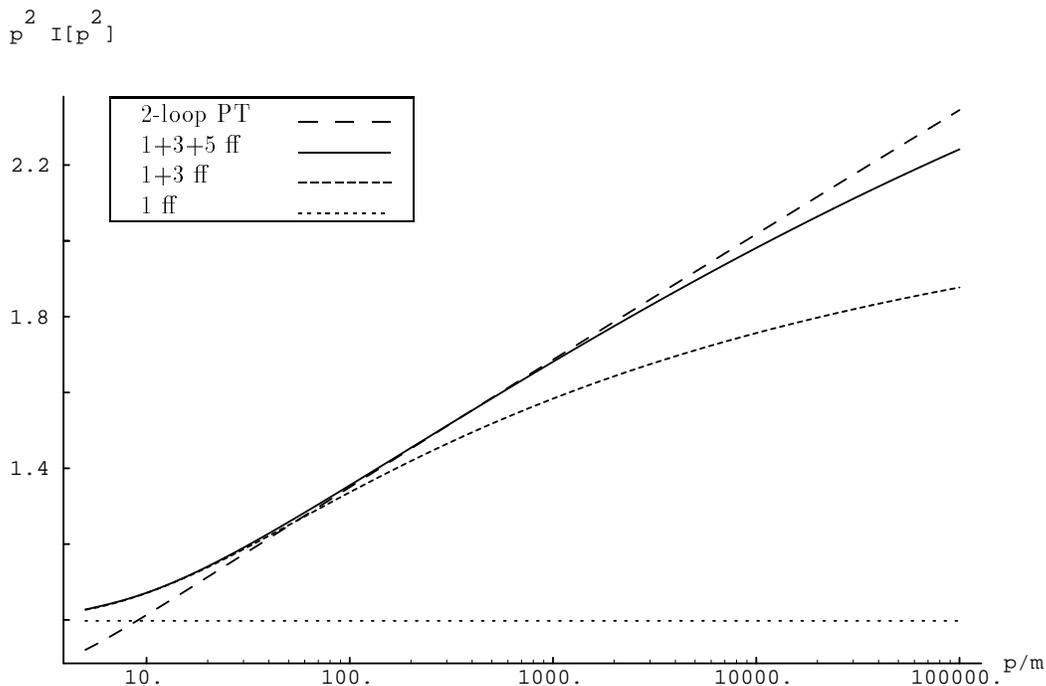}
%\vskip 10mm
\end{flushleft}
\caption{Spin two-point function. Comparison: Form factor approach 
versus 2-loop perturbation theory; logplot of $p^2 I(p^2)$ against $p/m$.
The normalization of the PT curve is fixed according to (\ref{lb1}).} 
\label{spin2}
\end{figure}
%%%%%%%%%%%%%%%%%%%%%%%%%%%%%%%%%%%%%%%%%

%%%%%%%%%%%%%%%%%%%%%%%%%%%%%%%%%%%%%%%
% figure curr2
\begin{figure}[htb]
\begin{flushleft}
%\vskip 10mm
\leavevmode
\epsfxsize=170mm
\epsfbox{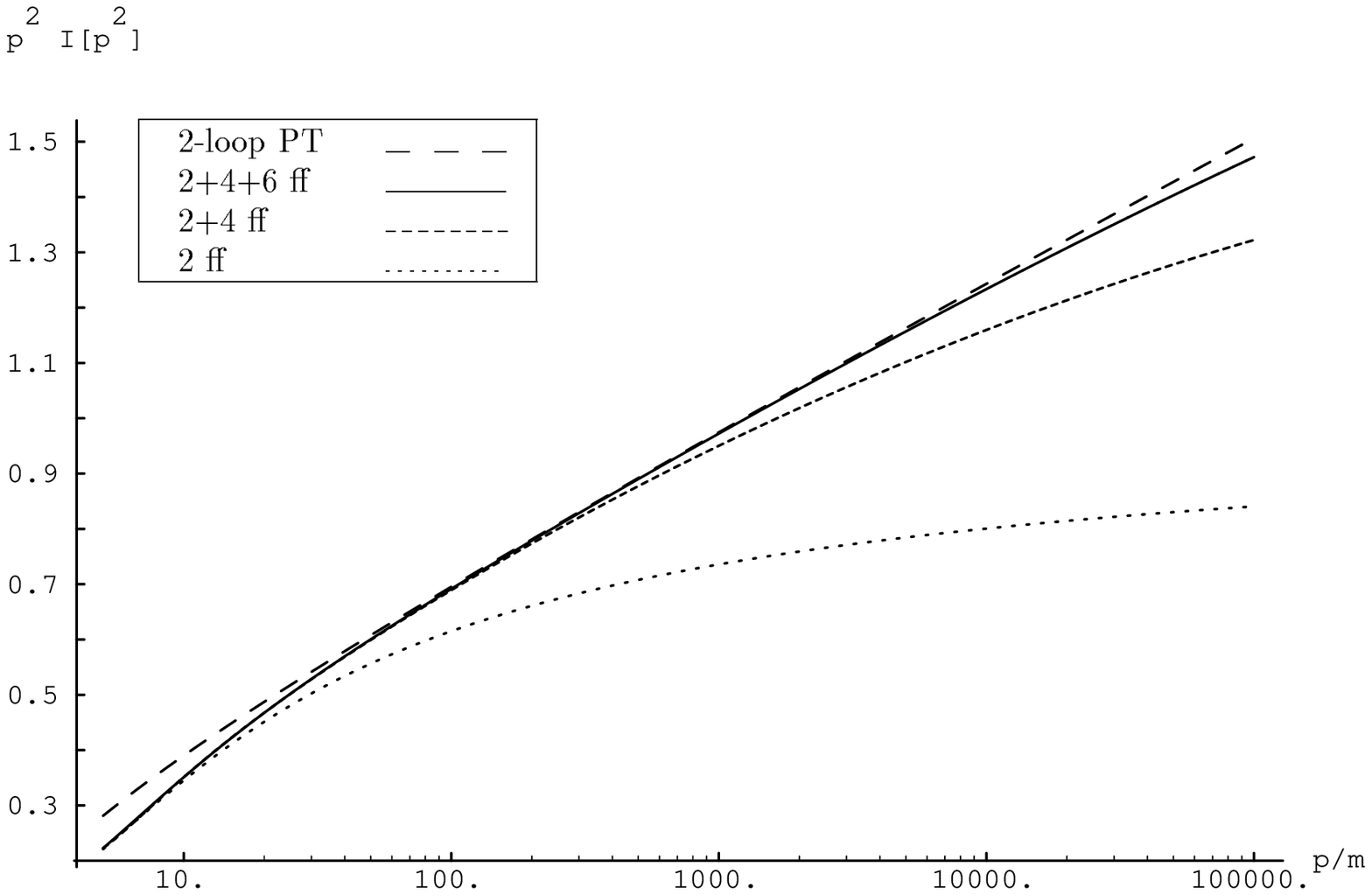}
%\vskip 10mm
\end{flushleft}
\caption{Current two-point function. Comparison: Form factor approach 
versus 2-loop perturbation theory; logplot of $p^2 I(p^2)$ against $p/m$.} 
\label{curr2}
\end{figure}
%%%%%%%%%%%%%%%%%%%%%%%%%%%%%%%%%%%%%%%%%

In Figures \ref{spin2} and \ref{curr2} again the Fourier transform 
$I(p^2)$ of the 2-point function of the Spin field and the Noether 
current are shown, computed once in 2-loop PT and once via (\ref{sp10})
by truncation of the form factor series. 
In the spin case the normalization of the PT curve is fixed according 
to (\ref{lb1}), so that as in the current case no free parameter enters 
the comparison. Let us define the perturbative
regime to be the energy range for which the 2-loop PT and the truncated 
form factor curve coincide within $1\%$. In both the Spin and the Current
case a large perturbative regime is found to exist. Having fixed 
$m/\Lambda$ and $\lambda_1$ no adjustable parameters enter the PT results.
The very existence of such a perturbative regime thus supports the proposed
value of $m/\Lambda$ as well as the exact value of the normalization 
constant (\ref{lb1}). In the Spin case the 2-loop perturbative regime
is about $50 \lsim \mu/m \lsim 5000$ and in the current case it 
is about $100 \lsim \mu/m \lsim 2\cdot 10^4$. The lower bound of this interval
will remain unaffected when higher particle form factor contributions
are taken into account and thus is a genuine feature of the O$(3)$ model.
Notice however that the onset of the 2-loop perturbative regime 
occurs at much higher energies ($50 -100$ times the mass gap) than is 
sometimes pretended in the 4-dimensional counterpart of this situation. 
Probably this should be taken as a warning in 4-dimensional 
gauge theories. 

The upper bounds $\mu \approx 5000$ for the $(1+3+5)$-particle Spin
curves and $\mu \approx 10^4$ for the $(2+4+6)$-particle Current curve
are fully consistent with the intrinsic estimate for the validity of the
truncated form factor results given at the beginning of this section. 
These upper bounds of the perturbative regime, of course, are expected to 
move up as higher particle form factor contributions are taken into account. 
Indeed, the hypothesis of asymptotic freedom is that they approach 
infinity, so that the entire high energy regime is perturbative. 
As emphasized in the introduction, the issue of asymptotic freedom
can only be addressed if the high energy properties of the theory can 
be explored by means of a reliable technique {\em independent} of PT. 
Guided by the form factor results we formulate in the next section 
a conjecture about the structure of the $n$-particle spectral densities
for general $n$. Assuming the validity of this conjecture the extreme 
UV behavior of the two-point functions can be computed. 

%%%%%%%%%%%%%%%%%%%%%%%%%%%%%%%%%%%%%%%%%%%%%%%%%%%%%%%%%%%%%%%%%%%%%%%%
\newpage
\newsection{Scaling Hypothesis for Spectral Densities} 
\setcounter{totalnumber}{2}
\renewcommand{\textfraction}{0.01}
\renewcommand{\bottomfraction}{0.99}
\floatsep 0mm

The $n$-particle contributions to the spectral densities appear
to follow a remarkable `self-similarity' pattern. Suppose the scalarized
$n$-particle spectral density of one of the four local operators
considered to be given. Can one -- without actually doing the 
computation (which for large $k$ is practically impossible)
-- anticipate the structure of some $k >n$ particle spectral
density? We propose that the answer is affirmative and for large $n$ 
and $\lb >1$ is given by
\be
\rho^{(\lb n)}(\mu) \approx \frac{m}{\mu \lb^{\gamma}}\,
(\mu/m)^{1/{\lb^{1+\alpha}}} \;
\rho^{(n)}\left(m(\mu/m)^{1/{\lb^{1+\alpha}}}\right)\;,
\label{scal1}
\ee
where $m$ is the mass gap and the exponents $\gamma$ and $\alpha$ are 
given below. Taking $\lb = \frac{n+2}{n}$ for example yields a candidate
for the $(n+2)$-particle spectral density.
In the following we shall first give a precise formulation 
of the scaling law (\ref{scal1}) and present evidence for it. In section 
5.2 we then promote (\ref{scal1}) to a working hypothesis 
and explore its consequences. The results obtained have an interesting 
interplay with both PT and non-perturbative MC data. All pieces of 
information are found to be consistent, which we interpret as further 
supporting the hypothesis.    

\newsubsection{Evidence for Self-similarity and Scaling}  

Since the leading asymptotics of the spectral densities is given
by $1/\mu$ it is useful to consider $\mu\rho^{(n)}(\mu)$ instead 
of $\rho^{(n)}(\mu)$. Further a logarithmic scale is convenient
so that we are lead to introduce
\be
R^{(n)}_l(x) := m e^x \,\rho^{(n)}_l(m e^x)\;,\sspace 
\ln n \leq x < \infty \;,\;\;\; l=0,1\;.
\label{scal2}
\ee
Here $l=0,1$ as before correspond to the EM tensor \& TC density and 
Spin \& Current series, respectively. The graphs of these functions 
are again roughly `bell-shaped': Starting from zero at $x = \ln n$
they are strictly increasing,  reach a single maximum at some 
$x =\xi^{(n)}_l > \ln n$ and then decrease monotonically for all 
$x > \xi^{(n)}_l$. The position $\xi^{(n)}_l$ of the maximum
and its value $M^{(n)}_l := R^{(n)}_l(\xi^{(n)}_l)$ are two important
characteristics of the function, and hence of the spectral density.  
For $n\leq 6$ these data are collected in Table 1.

%%%%%%%%%%%%%%%%%%%%%%
\begin{table}{Table 1: Positions and values of the maxima}
\begin{center}
\begin{tabular}[t]{r|c c|c c}
n &$\;\;\;\;\xi^{(n)}_0\;\;$ & $\;\;10^4 M^{(n)}_0\;\;$
  &$\;\;\;\;\xi^{(n)}_1\;\;$ & $\;\;10^2 M^{(n)}_1\;\;$ 
\\[0.5ex] \hline 
2 & 0.953        & 317.3           & 1.155        & 27.33   \\
3 & 2.726        & 27.95           & 3.613        & 10.76   \\
4 & 4.720        & 7.870           & 6.631        & 6.736   \\
5 & 6.945        & 3.134           & 10.06        & 4.911   \\
6 & 9.344        & 1.517           & 13.79        & 3.867   
\end{tabular}
\end{center}
\end{table}
%%%%%%%%%%%%%%%%%%%%%%%%

The content of the scaling law (\ref{scal1}) is most transparent if the 
ordinate and abscissa of the graph are rescaled such that both
the value and the position of the maximum are normalized to unity. Thus 
define
\be
Y^{(n)}_l(z) := \frac{1}{M^{(n)}_l}\,R^{(n)}_l(\xi^{(n)}_l z)\;,
\sspace (\ln n)/\xi^{(n)}_l \leq z < \infty\;,\;\;\; l=0,1\;.
\label{scal3}
\ee
In order to have a common domain of definition we set $Y^{(n)}_l(z) 
= 0$ for $0\leq z\leq (\ln n)/\xi^{(n)}_l$. 
The proposed behavior of the spectral densities is as follows:
\vspace{5mm}

\noindent {\bf Scaling Hypothesis:}
\begin{itemize}
\item[(a)] {\em (Self-similarity) The functions $Y_l^{(n)}(z),\;n\geq 2$ 
converge pointwise to a bounded function $Y_l(z)$. The sequence of
$k$-th moments converges to the $k$-th moments of $Y_l(z)$ for $k+l=0,1$, 
i.e.}
\ba 
&&\lim_{n\ra \infty} Y^{(n)}_l(z) = Y_l(z)\;,\;\;\; z\geq 0\;,\nonum
&& \lim_{n\ra \infty}\int_0^{\infty}dz\, z^k Y_l^{(n)}(z) =
\int_0^{\infty}dz\, z^k Y_l(z)\;.
\label{scala}
\ea
\item[(b)] {\em (Asymptotic scaling) The parameters $\xi^{(n)}_l$ and 
$M^{(n)}_l$ scale asymptotically according to powers of $n$, i.e.}
\be
\xi_l^{(n)} \sim \xi_l\,n^{1+ \alpha_l}\;,\sspace
M_l^{(n)} \sim M_l\,n^{-\gamma_l}\;. 
\label{scalb}
\ee
\end{itemize}
\vspace{2mm}

\begin{figure}[t]
\begin{flushleft}
%\vskip -10mm
\leavevmode
\epsfxsize=170mm
\epsfysize=80mm
\epsfbox{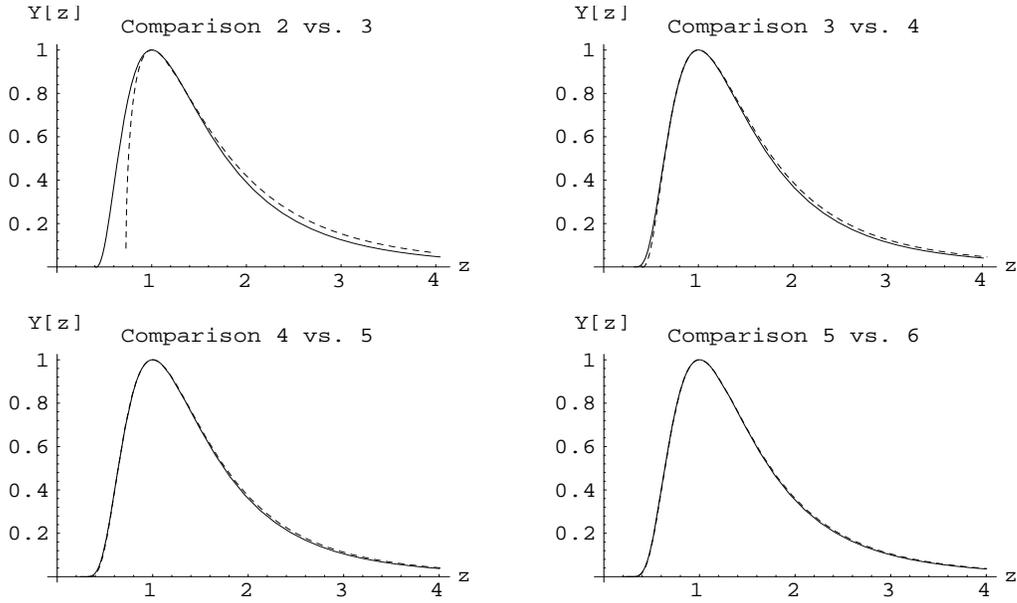}
\vskip -4mm
\end{flushleft}
\caption{Illustration of the self-similarity property of the rescaled $l=0$ 
spectral densities. The plots show $Y_0^{(n)}(z)$ (dashed) 
compared with $Y_0^{(n+1)}(z)$ (solid) for $n=2,3,4,5$.}
\label{SSem}
\end{figure}

\begin{figure}[bh]
\begin{flushleft}
\vskip -10mm
\leavevmode
\epsfxsize=170mm
\epsfysize=80mm
\epsfbox{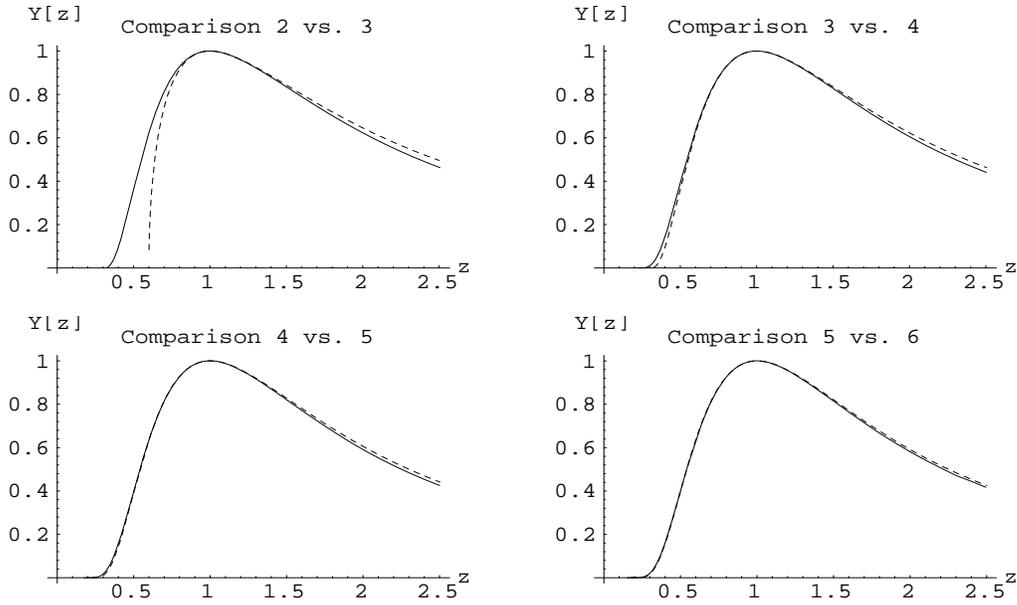}
\vskip -4mm
\end{flushleft}
\caption{Illustration of the self-similarity property of the rescaled $l=1$ 
spectral densities. The plots show $Y_1^{(n)}(z)$ (dashed) 
compared with $Y_1^{(n+1)}(z)$ (solid) for $n=2,3,4,5$.}
\label{SScurr}
\end{figure}

Note that the convergence in (a) is weaker than uniform convergence. 
Feature (a) in particular means that for sufficiently large $n$ the 
graphs of two subsequent members $Y^{(n-1)}_l(z)$ and $Y^{(n)}_l(z)$
should become practically indistinguishable. This appears to be 
satisfied remarkably well even for small $n=4,5,6$, as is 
illustrated in Figures \ref{SSem} and \ref{SScurr}.

%%%%%%%%%%%%%%%%%%%%%%%%%%%%%%%%%%%%%%%
% figure SSem
%%%%%%%%%%%%%%%%%%%%%%%%%%%%%%%%%%%%%%%%%
% figure SScurr
%%%%%%%%%%%%%%%%%%%%%%%%%%%%%%%%%%%%%%%%%

To substantiate proposal (b) let us prepare some further characteristics 
of the functions $R_l^{(n)}(x),\;l=0,1$: Its integral, its first, respectively 
minus first moment, and the strength of the asymptotic decay.   
In formulae
\ba
\begin{array}{ll}
c_0^{(n)} = {\displaystyle 12\pi \int_0^{\infty} dx  R_0^{(n)}(x)} \bspace &
c_1^{(n)} = {\displaystyle \int_0^{\infty} dx R^{(n)}_1(x)} \;, 
\\[1.5ex]
h^{(n)}_0 = {\displaystyle \int_0^{\infty} dx\,x R^{(n)}_0(x)}\;, & 
h^{(n)}_1 = {\displaystyle \int_0^{\infty} \frac{dx}{x} R^{(n)}_1(x)}
\\[1.5ex]
R^{(n)}_0(x) \sim 
{\displaystyle \frac{A^{(n)}_0}{x^4}}\;,\;\;x \ra \infty &
R^{(n)}_1(x) \sim 
{\displaystyle \frac{A^{(n)}_1}{x^2}}\;,\;\;x \ra \infty \;.
\end{array}
\label{scalR}
\ea 
In practice of course one will evaluate the constants $c^{(n)}_l$ and 
$h^{(n)}_l$ directly as $(n-1)$-fold unconstrained integrals, 
as in (\ref{asy3}), rather
than first computing the spectral densities and then evaluating their 
moments. For $n\leq 6$ the results for these constants are listed in 
Table 2.  

%%%%%%%%%%%%%%%%%%%%%%%%%%%%
\begin{table}[ht]{Table 2: Integrals, $\pm$ first moments and 
decay constants}
\begin{center}
\begin{tabular}[ht]{r|c c c|c c c}
n &$\;\;\;\;c_0^{(n)}\;\;$ & $\;\;10^2 h^{(n)}_0\;\;$ & $\;\;A^{(n)}_0\;\;$ 
  &$\;\;\;\;c_1^{(n)}\;\;$ & $\;\;10 h^{(n)}_1\;\;$ & $\;\;A^{(n)}_1\;\;$ 
\\[0.5ex] \hline 
2 & 1.6027  &  8.231    & 0.727   &  1.009  &  3.958  & 1.938   \\
3 & 0.4104  &  5.251    & 1.866   &  1.140  &  1.693  & 4.977   \\
4 & 0.1943  &  4.140    & 3.308   &  1.242  &  1.054  & 8.821   \\
5 & 0.1117  &  3.430    & 5.006   &  1.327  &  0.761  & 13.35  \\
6 & 0.07185 &  2.930    & 6.938   &  1.400  &  0.594  & 18.50 
\end{tabular}
\end{center}
\end{table}
%%%%%%%%%%%%%%%%%%%%%%%%%%%%%%%%%%%
In addition one has $c_1^{(1)} = \pi/4$. The decay constants $A_l^{(n)}$
are difficult to compute directly since the truly asymptotic behavior
of the spectral densities sets in only at astronomically large energies
$\mu \sim m\,e^{10\,n^{5/4}}$. Using the exact relations (\ref{asy4}) they 
can however accurately be computed from the integrals $c_1^{(n)}$.    

\begin{figure}[htb]
\begin{flushleft}
%\vskip 10mm
\leavevmode
\epsfxsize=170mm
\epsfbox{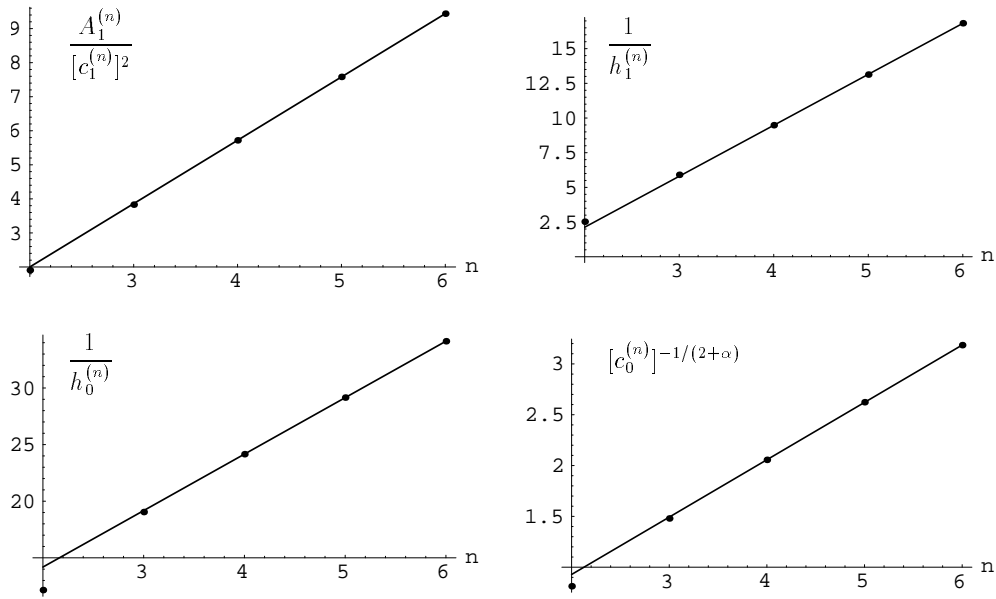}
%\vskip 10mm
\end{flushleft}
\caption{Two-parameter linear fits for the $n$-dependence of 
various parameter combinations.}
\label{fits}
\end{figure}

Making use of the convergence (\ref{scala}) one can deduce scaling laws 
for the quantities (\ref{scalR}). To this end we introduce the 
corresponding quantities for the universal shape functions $Y_l(z)$
\ba
\begin{array}{ll}
c_0^* = {\displaystyle 12\pi \int_0^{\infty} dz  Y_0(z)\;,} \bspace &
c_1^* = {\displaystyle \int_0^{\infty} dz Y_1(z)} \;, 
\\[1.5ex]
h^*_0 = {\displaystyle \int_0^{\infty} dz\,z Y_0(z)}\;, & 
h^*_1 = {\displaystyle \int_0^{\infty} \frac{dz}{z} Y_1(z)\;,}
\\[1.5ex]
Y_0(z) \sim 
{\displaystyle \frac{A^*_0}{z^4}}\;,\;\;z \ra \infty\;, &
Y_1(z) \sim 
{\displaystyle \frac{A^*_1}{z^2}}\;,\;\;z \ra \infty \;.
\end{array}
\label{scalY}
\ea 
For large enough $n$ we can approximate $R_l^{(n)}(x)$ pointwise and
w.r.t. the moments by 
\be
R^{(n)}_l(x) \approx \frac{M_l}{n^{\gamma_l}}\;
Y_l\left(\frac{x}{\xi_l n^{1+\alpha_l}}\right)\;.
\label{scal4}
\ee
Combining (\ref{scalb}) - (\ref{scal4}) one finds
\ba
\begin{array}{rclcrcl}
c^{(n)}_0 & \sim &  c_0^* M_0 \xi_0 \;n^{1+\alpha_0 -\gamma_0}\;, 
&\sspace 
c^{(n)}_1 & \sim &  c_1^* M_1 \xi_1 \;n^{1+\alpha_1 -\gamma_1}\;,
\\[1.5ex] 
h_0^{(n)} & \sim & h_0^* M_0 \xi_0^2 \; n^{2+2\alpha_0 -\gamma_0}\;,
&\sspace
h_1^{(n)} & \sim & h_1^* M_1 \;n^{-\gamma_1}\;,\sspace
\\[1.5ex]
A_0^{(n)} & \sim & A_0^* M_0 \xi_0^4\; n^{4 + 4\alpha_0 -\gamma_0}\;,
&\sspace 
A_1^{(n)} & \sim & A_1^* M_1 \xi_1^2\; n^{2 +2\alpha_1 -\gamma_1}\;.
\end{array}
\label{scal5}
\ea 
Two important constraints arise from the non-linear relations (\ref{asy4})
for the asymptotic decay constants $A_l^{(n)}$. Combined with 
(\ref{scal5}) this yields the following relations among the parameters 
\begin{subeqnarray}  
& \gamma_1 = 1\;, \sspace  &
\frac{A_1^*}{(c_1^*)^2 M_1} = \pi\, 
\frac{\Gamma(\alpha_1 +1)^2}{\Gamma(2\alpha_1 +2)}\;,\\ 
&\gamma_0 = 3 + 4\alpha_0 -2\alpha_1\;,\sspace &
\;A_0^* M_0 \xi_0^4 = \frac{3}{8}A_1^* M_1 \xi_1^2\;.
\label{scal6}
\end{subeqnarray}
(\ref{scal6}b) follows trivially from $A_1^{(n)} = \frac{3}{8}
A_0^{(n)}$. To arrive at (\ref{scal6}a) we approximated 
$$
\sum_{k=1}^{n-1}[k(n-k)]^{\mu} \approx n^{2\mu +1} 
\int_0^1 dy [y(1-y)]^{\mu}\;,
$$
which is valid for large $n$.
The exponents $\gamma_l$ are determined in terms of 
$\alpha_0$ and $\alpha_1$, which will later turn out coincide
\be
\alpha_0 = \alpha_1 =: \alpha\;.
\label{scal7}
\ee
Re-inserting (\ref{scal6}) and (\ref{scal7}) 
into (\ref{scal5}) one arrives at a simplified set of scaling relations. 
In particular
\ba
\frac{1}{h^{(n)}_1} \,\sim\,\frac{1}{h_1}\,n\;,
&\sspace & h_1 := h_1^*M_1\;,\nonum
\frac{1}{h^{(n)}_0}\,\sim\,\frac{1}{h_0}\,n\;,
&\sspace & h_0 := h_0^*M_0\xi_0^2\;,\nonum
[c_0^{(n)}]^{-1/(2+\alpha)} \,\sim \, c_0 \,n\;, 
&\sspace & c_0 := (c_0^* M_0 \xi_0)^{-1/(2+\alpha)}\;,
\label{scal8}
\ea
are predicted to scale linearly with $n$. The same holds for the 
following two ratios, which provide a convenient way to determine 
the remaining exponent $\alpha$ from the slope of the linear
function
\be
\frac{A^{(n)}_1}{[c_1^{(n)}]^2} = 
\frac{8}{3}\,\frac{A^{(n)}_0}{[c_1^{(n)}]^2} \sim
\pi \frac{\Gamma(\alpha +1)^2}{\Gamma(2\alpha +2)}\;n\;.
\label{scal9}
\ee
The predicted linear scaling in (\ref{scal8}), (\ref{scal9}) appears 
to be satisfied fairly well even for small $n=4,5,6$ as 
is illustrated in Figure \ref{fits}. 

%%%%%%%%%%%%%%%%%%%%%%%%%%%%%%%%%%%%%%%
% figure fits
%%%%%%%%%%%%%%%%%%%%%%%%%%%%%%%%%%%%%%%%%

The fits are always two-parameter
linear fits (slope and ordinate) on the largest three data points
$n=4,5,6$. Using (\ref{scal9}) the exponent $\alpha$ comes out to be 
\be
\alpha \approx 0.273\;.
\label{alpha}
\ee
Using this value in the fitting of $[c_0^{(n)}]^{-1/(2+\alpha)}$ the
predicted linear scaling is confirmed. The quality of the $1/h^{(n)}_0$
fit confirms (\ref{scal7}). For later use let us also note the 
values of the slopes in (\ref{scal8})
\be
1/h_1 = 3.674\;,\;\;\;1/h_0 = 4.988\;,\;\;\; c_0 = 0.5643 \;.   
\label{slopes}
\ee
Similar fits can be made for the values and the positions of the maxima, 
given numerically in Table 1. They are similarly 
convincing; we refrain from displaying them.

%%%%%%%%%%%%%%%%%%%%%%%%%%%%%%%%%%%%%%%%%%%%%%%%%%%%%%%%%%%%%%%%%%%%%
\newsubsection{Summation of $n$-particle contributions}

Having at hand candidate expressions for the $n$-particle spectral 
densities as given in (\ref{scal4}) one can evaluate their sum.
Suppose that for $n< n_0$ the $n$-particle spectral densities 
have been computed explicitly; in our case $n_0 =7$. Decomposing
the sum
\be
R_l(x) = \sum_{n=2-l}^{n_0-1} R_l^{(n)}(x) + 
\sum_{n=n_0}^{\infty} R_l^{(n)}(x) \;,
\label{scal11}
\ee
we wish to evaluate the second term. Using (\ref{scal4}) the 
sum can be expressed in terms of a suitable integral transform 
of the shape function $Y_l(z)$. Here we shall need only the large 
$x$ asymptotics of the sum, in which case it is sufficient to 
approximate the sum over $n$ by an integral. One finds for 
the large $x$ behavior
\be
\sum_{n=n_0}^{\infty} R_l^{(n)}(x) \rra \frac{M_l}{1+\alpha_l}\, 
\left( \frac{x}{\xi_l} \right)^{\frac{1-\gamma_l}{1+\alpha_l}}\;
\int_0^{\infty} \frac{dz}{z}\; z^{\frac{\gamma_l-1}{1+\alpha_l}}\; 
Y_l(z)\;,\sspace x\ra \infty\;,
\label{scal12}
\ee
where the corrections are at least one power down in $x$.
Clearly the physical spectral densities can be evaluated 
similarly; summing only over odd/even particle numbers will simply 
result in an extra factor $1/2$ in the asymptotic expression (\ref{scal12}).
In particular in the isospin $1$ case the exponent $\gamma_1 =1$ 
yields the asymptotics $R_1(x) \ra A_1$. One can show that the 
asymptotics of $R_0(x)$ must always be down by two orders in $x$,
compared to that of $R_1(x)$. This is related to, although not a direct 
consequence of (\ref{asy2}). One concludes that the $x \ra \infty$   
asymptotics in the isospin zero case must be $R_0(x) \ra A_0/x^2$. Comparing
with (\ref{scal12}) this requires $\alpha_0 = \alpha_1$, as claimed
before. Moreover one obtains the relations 
\begin{subeqnarray}
A^{\rm curr} &=& \frac{\pi}{4} A^{\rm spin} = \frac{h_1}{2(1+\alpha)}\;,
\\
A^{\rm EM} &=& 4 A^{\rm top} = \frac{h_0}{2(1+\alpha)}\;,
\label{scal13}
\end{subeqnarray}
which have various interesting consequences. (\ref{scal13}a)
relates $A^{\rm curr}$ to $A^{\rm spin}$. This is interesting because 
$A^{\rm curr}$ is accessible to perturbation theory, while 
$A^{\rm spin}$ is some previously unknown non-perturbative constant, 
which via (\ref{o311}), (\ref{o313})
determines the short-distance normalization of the Spin two-point function. 
On the other hand both constants also get expressed in terms of the
scaling hypothesis. Using (\ref{alpha}) and  (\ref{slopes}) one finds
$(A^{\rm curr})_{SH} = 1/9.3553 =1.007/3\pi$. Since perturbation theory gives 
$(A^{\rm curr})_{PT}= 1/3\pi$ one sees that not only the functional
form of the asymptotic behavior obtained from the resummation of the spectral
resolution agrees with that predicted by PT, but also the numerical value 
of the leading coefficient agrees with an accuracy better than $1\%$.
Depending on the viewpoint one can interpret
this as supporting the validity of PT and/or our scaling hypothesis
\footnote{The extrapolation based on the scaling hypothesis contains
a (neglegible) numerical error as well as a systematic error. The 
latter is due to the uncertainty introduced by doing the fits for 
moderately large particle numbers n=4,5,6. In principle there may also
be subleading (not powerlike) terms in the scaling laws (\ref{scalb}).
A rough feeling for the size of the systematic errors can be obtained
by ad-hoc changing $\alpha$ in (\ref{alpha}) to $\alpha =0.25$. One 
finds $(A^{\rm curr})_{SH} = 1.026/3\pi;\;(A^{\rm EM})_{SH} =
1.008/4\pi; \;c_{SH} = 1.998$.}. 
Accepting the PT value for $A^{\rm curr}$ we get the exact prediction
for the non-perturbative constant $\lb_1 = A^{\rm spin}$ anticipated in 
(\ref{lb1})
\be
\lb_1= A^{\rm spin} = \frac{4}{3\pi^2} = 0.135095\;.
\label{scal14}
\ee
On the other hand $A^{\rm spin}$ has been evaluated by Monte Carlo 
simulations \cite{PSMC} with the result $(A^{\rm spin})_{MC} = 0.135(2)$, 
convincingly close to the proposed exact result. The similar behavior of 
the Spin and Current two-point functions and spectral densities for 
large energy is again an O$(3)$ speciality, as can be seen from
(\ref{o314}).

A similar pattern underlies (\ref{scal13}b). Now both $A^{\rm EM}$
and $A^{\rm top}$ can be computed in PT. The results are given in
(\ref{o3top6}), (\ref{o3EM7}) and yield in particular 
$(A^{\rm EM})_{PT} = 4 (A^{\rm top})_{PT}$.  
The non-perturbative result now is that this transfers to relations 
for the exact spectral densities at {\em all} energy scales
\be
\rho^{\rm EM}(\mu) = \sum_{k=1}^{\infty} \rho_0^{(2k)}(\mu)\,,\sspace 
\rho^{\rm top}(\mu) = \frac{1}{4} \sum_{k=1}^{\infty} 
\rho_0^{(2k+1)}(\mu)\;, 
\label{scal15}
\ee
which is the announced result (\ref{lambda0}). Consistency then 
requires that the numerical value for $A^{EM}$ computed via the 
scaling hypothesis coincides with the perturbative value.
Combining (\ref{alpha}), (\ref{slopes}) and (\ref{o3EM7}) one finds 
\be
(A^{\rm EM})_{SH} = 1/12.700 =0.989/4\pi
\;,\sspace (A^{\rm EM})_{PT} =1/4\pi\;.
\ee
Note that here the matching between full and perturbative 
dynamics already concerns a 1-loop coefficient and is accurate to 
$1\%$. Again this cuts both ways, supporting the validity of PT and/or 
our scaling hypothesis and thus also the proposed normalization 
constant $\lb_0 =1/4$ of the topological charge density operator. 
 
Finally consider the central charge as defined in (\ref{cdef}).
In the spectral resolution it again decomposes into a sum of 
$n$-particle contribution. The first few terms are given in Table~2, i.e.
\be
c^{(2)}_0 = 1.603\;,\;\;c^{(4)}_0 =0.194\;,\;\;c^{(6)}_0 = 0.072\;,\sspace
c^{(n)}_0 = 12\pi \int_0^{\infty}d\mu \,\rho^{(n)}_0(\mu)\;.
\ee
The remainder of the series can be evaluated by means of the scaling
hypothesis. Using (\ref{slopes}), (\ref{alpha}) one finds 
\be
c_{SH} = c^{(2)}_0 + c^{(4)}_0 + c^{(6)}_0 + \sum_{k=4}^{\infty}
(2k c_0 - 0.200)^{-(2+\alpha)} = 1.997\;.
\ee 
Here we also used the ordinate of the linear fit.
(Tree level)
PT predicts $c_{PT}=2$, corresponding to the two unconstrained bosonic
degrees of freedom. The form factor computation shows that this is 
compatible with the non-perturbative low energy dynamics of the model
and provides further support for the scaling hypothesis (see the
footnote on page 38). An alternative 
non-perturbative consistency check is provided by the thermodynamic 
Bethe Ansatz \cite{FZ} and also yields $c=2$. Let us point out that the 
value $c=2$ does not carry much information about the nature of the UV 
limiting CFT. For a number of reasons it cannot be the Gaussian $c=2$ 
model.

%%%%%%%%%%%%%%%%%%%%%%%%%%%%%%%%%%%%%%%%%%%%%%%%%%%%%%%%%%%%%%%%%%%%%%
\newsubsection{Towards the exact two-point functions}

The scaling hypothesis also allows one to numerically compute 
the two-point functions at {\em all} energy/length scales. 
In coordinate space this simply amounts to performing the 
integral in the spectral representation, once the full 
spectral density is available via (\ref{scal11}) and (\ref{scal4}). In 
momentum space one can combine eq.~(\ref{sp10}) and (\ref{scal4})
to evaluate the sum over all $n$-particle contributions to the 
functions $I(p)$, basically giving the Fourier transform of the 
two-point function. The results will be described in \cite{BN1}.
 
%%%%%%%%%%%%%%%%%%%%%%%%%%%%%%%%%%%%%%%%%%%%%%%%%%%%%%%%%%%%%%%%%%%%%%
\newsection{Conclusions}

In this paper we studied the two-point functions of the four
physically most interesting operators in the O$(3)$ NLS model. We have
chosen the O$(3)$ model because it is the simplest 2-dimensional
theory that resembles in many aspects QCD. Concerning the on-shell
features various exact results, in particular the exact two-particle
S-matrix are known \cite{Zam,Luscher}, due to its remarkable 
integrability properties. On the other hand there exist a large number 
of MC studies of the model \cite{LW,UWolff,LWW,PSMC,MC}, not making 
use of this integrability. Here we employed the form factor bootstrap
method for integrable QFTs to compute off-shell features that can be
compared both with PT and with MC data. A simplifying feature of
the O$(3)$ model (as compared, say, to the higher O$(N)$ models) is 
that the computation of the form factors is relatively straightforward 
due to their essentially polynomial nature. Once the form factors are 
known, the two-point functions can be evaluated in terms of a spectral 
resolution; that is as an infinite sum where multiple integrals of the 
squares of the $n$-particle form factors enter, the sum running over all 
particle numbers $n$. In practice one has to truncate this infinite sum 
after the first few terms. In models with a diagonal S-matrix and a 
power-like approach of the UV limit, application of this technique showed 
extremely fast convergence of the truncated sum \cite{YZ,Conv1,Conv2}.%
\footnote{Nevertheless we expect the convergence still to be non-uniform
in the sense that the UV behavior of the infinite sum is different 
from that of each partial sum.}
In an asymptotically free theory one expects less fast convergence, because
the approach to the asymptotic form is only logarithmic. Thus, in order
to really check the validity of PT (which has been questioned 
\cite{PS1,PS2}) we needed to extrapolate our results beyond the largest 
particle number, presently 6, for which we carried out the computations 
explicitly. The extrapolation was based on a novel scaling hypothesis.
Using this hypothesis we have been able to compute the extreme UV properties
independent of PT. Remarkably we found very good agreement with PT: 
Within 1\% for a RG improved tree level coefficient in the case of the 
Current and also within 1\% for a RG improved 1-loop coefficient in the 
case of the EM tensor. This is very strong evidence for PT. 
The agreement between our extrapolated data and PT also indirectly
confirms the proposed exact value of the perturbative $\Lambda$
parameter \cite{HN}. On the basis of the evidence presented in section 5
we regard this scaling hypothesis as very plausible, although an
analytical proof remains to be found and would be highly desirable.
On the other hand, the nature of this hypothesis is completely independent 
of the assumptions underlying the derivation of the usual RG
improved expressions based on PT, so that the agreement of the results
is still remarkable.

A further result is the exact determination of two, previously unknown,
non-perturbative constants $\lambda_0$ and $\lambda_1$.
Using the values of these numbers, we can write the exact 3-particle
matrix element of the properly normalized topological charge density
operator $q(x)$ as
\be
\bra0|\, q(0)\,|a_3,\th_3;a_2,\th_2;a_1,\th_1\ket=
\frac{\pi^{\frac{7}{2}}}{4}\,
\Big[m^2-M^{(3)}(\th_3,\th_2,\th_1)^2\Big]\,
\Psi(\th_3,\th_2,\th_1)\,\epsilon^{a_3a_2a_1}\,,
\label{concl1}
\ee
where $M^{(3)}(\th_3,\th_2,\th_1)$ is the three-particle invariant
mass. Further the short distance expansion of the Spin two-point 
function is now unambiguously fixed to be
\be
S^{\rm spin}(x)=\frac{1}{3\pi^3}\left(\ln m r\right)^2+
O\left(\ln\ln m r\cdot\ln m r\right)\,,
\label{concl2}
\ee
where $m$ is the mass gap and $r =\sqrt{x_1^2 +x_2^2}$. 
Note that the values of the overall constants in (\ref{concl1})
and (\ref{concl2}) follow from the scaling hypothesis alone, we do
not need here the results of any numerical fitting procedure.

Finally we remark that our scaling hypothesis may be viewed as a
2-dimensional analogue of the KNO scaling \cite{KNO} in QCD.
We also expect that scaling hypotheses of a similar type can be 
formulated in many other 2-dimensional models, giving simultaneous access to 
their off-shell properties at all energy/length scales.

%%%%%%%%%%%%%%
\vspace{13mm}

{\tt Acknowledgements:} We wish to thank H. Lehmann, F. Niedermayer,
A. Patrascioiu, E. Seiler and P. Weisz for stimulating discussions. 
In particular we thank E. Seiler for initially suggesting to study the 
$n$-dependence of the spectral densities with hindsight to extrapolations. 
A.P. and E.S. we thank in addition for generously allowing us to use 
their data prior to publication. M.N. acknowledges support by the 
Reimar L\"ust fellowship of the Max-Planck-Society.  
J.B. wishes to thank the Theory Group at MPI Munich for their hospitality.

\vspace{23mm}

%%%%%%%%%%%%%%%%%%%%%%%%%%%%%%%%%%%%%%%%%%%%%%%%%%%%%%%%%%%%%%%%%%%%%%%

\setcounter{section}{0}
{\large \bf Appendix: List of form factor squares}
\renewcommand{\theequation}{A.\arabic{equation}}
\bigskip
\smallskip

Here we list the results for the form factor squares for 
the EM tensor \& TC density and  Current \& Spin series up to 
6 particles. The corresponding Mathematica files can be obtained 
from the authors upon request. The squares are boost
invariant symmetric polynomials in the rapidities and therefore
conveniently described in terms of a basis in this space of polynomials. 

Let $P^{(n)}(N,p)$ denote the space of homogeneous symmetric polynomials  
in $n$ variables that are of total degree $N$ and partial degree 
$p$. By partial degree we mean the power with which 
an individual variable enters. A basis for  $P^{(n)}(N,p)$ can be  
obtained as follows. For fixed $n$ let $\sigma_1^{(n)},\ldots,\sigma_n^{(n)}$  
denote the elementary symmetric polynomials in $\th_1,\ldots,\th_n$, i.e.
\be
\sigma_k^{(n)} =\sum_{i_1<\ldots <i_k}\th_{i_1}\ldots \th_{i_k}\;.
\ee
Let $\lb = (\lb_1,\ldots, \lb_p),\;\lb_1\leq \lb_2\leq \ldots\leq \lb_p$ 
be a partition of $N$ into $p$ parts less or equal to $n$,  
i.e. $\sum_i \lb_i = N,\;1\leq\lb_i \leq n,\;1\leq i\leq p$. Running through 
all those partitions, the assignment 
$$
(\lb_1,\ldots, \lb_p) \rra \sigma^{(n)}_{\lb_1}\ldots 
\sigma^{(n)}_{\lb_p}
$$
provides a basis of $P^{(n)}(N,p)$. However these functions are 
not boost invariant and one would like to have a description of the 
boost invariant subspace $P_{inv}^{(n)}(N,p)$ of $P^{(n)}(N,p)$. Boost 
invariance can be incorporated by switching to symmetric polynomials 
$\tau^{(n)}_k$ defined by
\ba
&& \tau^{(n)}_k(\th_1,\ldots,\th_n) = 
\sigma^{(n)}_k(\widehat{\th}_1,\ldots,\widehat{\th}_n)\;,
\sspace 2\leq k\leq n\;,\nonum
&& \tau^{(n)}_1 = \frac{1}{n}\sigma^{(n)}_1\;,\sspace
\widehat{\th}_j = \th_j -\frac{1}{n}(\th_1 + \ldots +\th_n)\;.
\ea
All monomials in the $\tau^{(n)}_k$'s ($k\geq2)$ are manifestly boost 
invariant. The price to pay is that the partial degree is no longer 
manifest. For a monomial in the $\sigma^{(n)}_k$'s the partial degree
is simply given by the number of
$\sigma^{(n)}_k$ factors. This is no longer the case for monomials 
in the $\tau^{(n)}_k$'s. In fact, inverting the relation
\be
\sigma_k = { n\choose k}\, \tau_1^k + \sum_{j=2}^k{n-j \choose k -j}\;
\tau_1^{k-j}\,\tau_j\;,\sspace k =1,\ldots, n
\ee
(here and henceforth the superscripts $(n)$ are suppressed) one
sees that $\tau_k =  (-\sigma_1)^k { n\choose k} (1-k)
n^{-k} +\ldots + \sigma_k$.
Thus for a monomial in the $\tau_k$'s the total and the partial degree
coincide. Linear combinations of such monomials however may have a 
partial degree that is less than their total degree. 

The general structure of the polynomials $G^{(n)}_l(\th)$
can now be described as follows. 
\be
G^{(n)}_l(\th) = G^{(n)}_{l;0}(\th) + 
\pi^2 G^{(n)}_{l;2}(\th) + \ldots + 
\pi^{N} G^{(n)}_{l;N}(\th)\;,\;\;\; G^{(n)}_{l;2k}(\th) 
\in P_{inv}^{(n)}(N-2 k,p)\;.
\ee
The overall leading terms $G^{(n)}_{l;0}(\th)$ are given in 
(\ref{overall}). Below we list the results for the polynomials 
$G^{(n)}_l(\th), \;n\leq 6$ for the Spin \& Current and 
EM tensor \& TC density series: 
\pagebreak[2]
\vspace{2mm}

Spin \& Current
\begin{eqnarray*}
&&G_1^{(1)}(\th) = 1\;,\sspace G_1^{(2)}(\th) = 2\;,\sspace
G_1^{(3)}(\th)  = 12[-\tau_2 + \pi^2]\;,\\[5mm]
&& G_1^{(4)}(\th) = -4[6\tau_2^3 + 9\tau_3^2 + 40 \tau_2\tau_4] + 
   8 \pi^2[25 \tau_2^2 + 44\tau_4] - 448 \pi^4 \tau_2 + 272 \pi^6\;,\\[5mm]
&&G_1^{(5)}(\th) = 
  [-48\tau_2^3\tau_3^2 - 72\tau_3^4 + 144\tau_2^4\tau_4 + 
  52\tau_2\tau_3^2\tau_4 + 352\tau_2^2\tau_4^2 - 640\tau_4^3 +
\\&& + 660\tau_2^2\tau_3\tau_5 + 3200\tau_3\tau_4\tau_5 + 
  4500\tau_2\tau_5^2]
\\[1mm]&&
 - 4 \pi^2[36\tau_2^5 - 64\tau_2^2\tau_3^2 + 692\tau_2^3\tau_4 + 
     295\tau_3^2\tau_4 + 80\tau_2\tau_4^2 +2925\tau_2\tau_3\tau_5 + 
  4375\tau_5^2] 
\\[1mm]&& 
 + 16 \pi^4[160\tau_2^4 + 97\tau_2\tau_3^2 + 909\tau_2^2\tau_4 - 
  20\tau_4^2 + 1950\tau_3\tau_5] 
\\[1mm]&& 
 - 16 \pi^6[978\tau_2^3 + 380\tau_3^2 + 1965\tau_2\tau_4] 
\\[1mm]&& 
 + 64 \pi^8[678\tau_2^2 + 395\tau_4] - 55120\, \pi^{10}\tau_2 + 
  24960\, \pi^{12}\;, \\[5mm] 
&& G_1^{(6)}(\th) = 4[-24\tau_2^3\tau_3^2\tau_4^2  
  - 36\tau_3^4\tau_4^2 + 72\tau_2^4\tau_4^3 + 26\tau_2\tau_3^2\tau_4^3 
  + 176\tau_2^2\tau_4^4 - 320\tau_4^5  
\\&& + 72\tau_2^3\tau_3^3\tau_5 + 108\tau_3^5\tau_5 -
     228\tau_2^4\tau_3\tau_4\tau_5 - 57\tau_2\tau_3^3\tau_4\tau_5 - 
     446\tau_2^2\tau_3\tau_4^2\tau_5 + 2208\tau_3\tau_4^3\tau_5  
\\&& + 108\tau_2^5\tau_5^2 - 801\tau_2^2\tau_3^2\tau_5^2 +
     1182\tau_2^3\tau_4\tau_5^2 - 3615\tau_3^2\tau_4\tau_5^2 - 
     430\tau_2\tau_4^2\tau_5^2 + 1800\tau_2\tau_3\tau_5^3  
\\&& + 5625\tau_5^4 - 180\tau_2^4\tau_3^2\tau_6 - 621\tau_2\tau_3^4\tau_6+
     480\tau_2^5\tau_4\tau_6 + 2082\tau_2^2\tau_3^2\tau_4\tau_6 - 
     2344\tau_2^3\tau_4^2\tau_6 
\\&& - 1710\tau_3^2\tau_4^2\tau_6 - 
    128\tau_2\tau_4^3\tau_6 + 3204\tau_2^3\tau_3\tau_5\tau_6 +
     3753\tau_3^3\tau_5\tau_6 - 744\tau_2\tau_3\tau_4\tau_5\tau_6 
\\&& + 14490\tau_2^2\tau_5^2\tau_6 - 34950\tau_4\tau_5^2\tau_6 - 
     288\tau_2^4\tau_6^2 + 13230\tau_2\tau_3^2\tau_6^2 -
     18696\tau_2^2\tau_4\tau_6^2 
\\&& + 40464\tau_4^2\tau_6^2 + 48870\tau_3\tau_5\tau_6^2 
     + 61992\tau_2\tau_6^3]
\\[1mm]&& 
   - 4\pi^2[72\tau_2^3\tau_3^4 + 108\tau_3^6 - 384\tau_2^4\tau_3^2\tau_4 - 
   408\tau_2\tau_3^4\tau_4 + 600\tau_2^5\tau_4^2 + 
   26\tau_2^2\tau_3^2\tau_4^2 
\\&& + 2248\tau_2^3\tau_4^3 + 1846\tau_3^2\tau_4^3 - 1504\tau_2\tau_4^4 
    - 264\tau_2^5\tau_3\tau_5 - 582\tau_2^2\tau_3^3\tau_5 
    - 3518\tau_2^3\tau_3\tau_4\tau_5  
\\&& - 6525\tau_3^3\tau_4\tau_5 + 
     8258\tau_2\tau_3\tau_4^2\tau_5 + 5574\tau_2^4\tau_5^2 -
     11685\tau_2\tau_3^2\tau_5^2 + 22220\tau_2^2\tau_4\tau_5^2  
\\&& - 14450\tau_4^2\tau_5^2 + 32625\tau_3\tau_5^3 + 1056\tau_2^6\tau_6 
     + 4110\tau_2^3\tau_3^2\tau_6 + 891\tau_3^4\tau_6 
     + 2768\tau_2^4\tau_4\tau_6  
\\&& +23766\tau_2\tau_3^2\tau_4\tau_6 - 54496\tau_2^2\tau_4^2\tau_6 + 
    30848\tau_4^3\tau_6 + 86190\tau_2^2\tau_3\tau_5\tau_6 - 
    54090\tau_3\tau_4\tau_5\tau_6  
\\&&
    + 78600\tau_2\tau_5^2\tau_6 - 47688\tau_2^3\tau_6^2 + 
    136404\tau_3^2\tau_6^2 +  51432\tau_2\tau_4\tau_6^2 + 378000\tau_6^3] 
\\[1mm]&& 
   + 16\pi^4[-126\tau_2^5\tau_3^2 - 24\tau_2^2\tau_3^4 + 336\tau_2^6\tau_4 - 
    1272\tau_2^3\tau_3^2\tau_4 - 1020\tau_3^4\tau_4 + 3818\tau_2^4\tau_4^2 
\\&& +3769\tau_2\tau_3^2\tau_4^2 + 426\tau_2^2\tau_4^3 + 376\tau_4^4 
     - 612\tau_2^4\tau_3\tau_5 - 6291\tau_2\tau_3^3\tau_5 
     + 6302\tau_2^2\tau_3\tau_4\tau_5  
\\&& +876\tau_3\tau_4^2\tau_5 + 21468\tau_2^3\tau_5^2 
     - 135\tau_3^2\tau_5^2 + 23185\tau_2\tau_4\tau_5^2 
     + 5640\tau_2^5\tau_6 + 29352\tau_2^2\tau_3^2\tau_6 
\\&& - 29108\tau_2^3\tau_4\tau_6 + 19149\tau_3^2\tau_4\tau_6 
     - 41452\tau_2\tau_4^2\tau_6 + 142074\tau_2\tau_3\tau_5\tau_6 + 
     33075\tau_5^2\tau_6 
\\&& - 93966\tau_2^2\tau_6^2 + 159882\tau_4\tau_6^2]  
\\[1mm]&& 
     - 8\pi^6[408\tau_2^7 - 4482\tau_2^4\tau_3^2 - 6564\tau_2\tau_3^4 + 
     18276\tau_2^5\tau_4 + 10016\tau_2^2\tau_3^2\tau_4 + 
     45484\tau_2^3\tau_4^2 
\\&& + 20301\tau_3^2\tau_4^2 - 5116\tau_2\tau_4^3 
     + 24628\tau_2^3\tau_3\tau_5 - 37755\tau_3^3\tau_5 + 
     117453\tau_2\tau_3\tau_4\tau_5 + 264035\tau_2^2\tau_5^2  
\\&& + 72675\tau_4\tau_5^2 + 69236\tau_2^4\tau_6 
     + 381147\tau_2\tau_3^2\tau_6 - 507872\tau_2^2\tau_4\tau_6 
     - 17628\tau_4^2\tau_6  
\\&& + 623655\tau_3\tau_5\tau_6 - 418404\tau_2\tau_6^2] 
\\[1mm]&& + 32\pi^8[2964\tau_2^6 - 3444\tau_2^3\tau_3^2 - 6138\tau_3^4 + 
     40900\tau_2^4\tau_4 + 30984\tau_2\tau_3^2\tau_4 +
     26211\tau_2^2\tau_4^2  
\\&& - 1048\tau_4^3 + 69483\tau_2^2\tau_3\tau_5 + 62973\tau_3\tau_4\tau_5 + 
     181365\tau_2\tau_5^2 + 33996\tau_2^3\tau_6  
\\&& + 182187\tau_3^2\tau_6 - 353994\tau_2\tau_4\tau_6 - 129717\tau_6^2]  
\\[1mm]&& - 32\pi^{10}[31348\tau_2^5 + 14205\tau_2^2\tau_3^2 
     + 177930\tau_2^3\tau_4 +  60772\tau_3^2\tau_4  
\\&& + 18686\tau_2\tau_4^2 + 237656\tau_2\tau_3\tau_5 + 184975\tau_5^2 - 
     6386\tau_2^2\tau_6 - 357082\tau_4\tau_6]  
\\[1mm]&& + 32\pi^{12}[168412\tau_2^4 + 78879\tau_2\tau_3^2 
     + 418856\tau_2^2\tau_4 - 8772\tau_4^2 + 260811\tau_3\tau_5 
     - 56784\tau_2\tau_6]  
\\[1mm]&& - 16\pi^{14}[1020460\tau_2^3 + 177855\tau_3^2 
     + 1047914\tau_2\tau_4 + 19362\tau_6]  
\\[1mm]&& + 128 \pi^{16}[220109\tau_2^2 + 70513\tau_4]  
      - 25589760\, \pi^{18}\tau_2+ 9265536\, \pi^{20}\;.  
\end{eqnarray*}
\vspace{5mm}

EM tensor \& TC density:
\begin{eqnarray*}
&& G_0^{(2)}(\th) = \frac{1}{\pi^2 -4\tau_2}\;,
\sspace G_0^{(3)}(\th) = 2 \\[5mm]
&& G_0^{(4)}(\th) = 4[12\tau_4 + \tau_2^2] - \pi^2 32\tau_2 
   + 28\,\pi^4\;,\\[5mm]
&& G_0^{(5)}(\th) = 4[2\tau_2^2\tau_3^2 - 6\tau_2^3\tau_4
   + 15\tau_3^2\tau_4 - 40\tau_2\tau_4^2 - 25\tau_2\tau_3\tau_5 
   - 625\tau_5^2] 
\\[1mm]&& + 8 \pi^2[3\tau_2^4 -11\tau_2\tau_3^2+73\tau_2^2\tau_4 +60\tau_4^2 
     +275\tau_3\tau_5] - 56 \pi^4[8\tau_2^3 + 5 \tau_3^2 + 40\tau_2\tau_4] 
\\[1mm]&& + 8 \pi^6 [253\tau_2^2 + 270\tau_4] - 3280 \pi^8 \tau_2 
   + 1680\, \pi^{10}\;,\\[5mm]
&& G_0^{(6)}(\th) = 4[4\tau_2^2\tau_3^2\tau_4^2 - 12\tau_2^3\tau_4^3 + 
   30\tau_3^2\tau_4^3 - 80\tau_2\tau_4^4 - 12\tau_2^2\tau_3^3\tau_5 + 
   38\tau_2^3\tau_3\tau_4\tau_5  
\\&& - 99\tau_3^3\tau_4\tau_5 + 246\tau_2\tau_3\tau_4^2\tau_5 - 
    18\tau_2^4\tau_5^2 + 165\tau_2\tau_3^2\tau_5^2 - 
    320\tau_2^2\tau_4\tau_5^2 - 450\tau_4^2\tau_5^2  
\\&& + 1125\tau_3\tau_5^3 + 30\tau_2^3\tau_3^2\tau_6 + 
    81\tau_3^4\tau_6 - 80\tau_2^4\tau_4\tau_6 - 
   234\tau_2\tau_3^2\tau_4\tau_6 + 212\tau_2^2\tau_4^2\tau_6  
\\&& - 48\tau_4^3\tau_6 - 438\tau_2^2\tau_3\tau_5\tau_6 +
     1530\tau_3\tau_4\tau_5\tau_6 - 9300\tau_2\tau_5^2\tau_6 - 
    48\tau_2^3\tau_6^2 - 5022\tau_3^2\tau_6^2  
\\&& + 10332\tau_2\tau_4\tau_6^2 - 59940\tau_6^3] 
\\[1mm]&&  + 8 \pi^2[6\tau_2^2\tau_3^4 - 32\tau_2^3\tau_3^2\tau_4 
     + 36\tau_3^4\tau_4 + 50\tau_2^4\tau_4^2 
     - 197\tau_2\tau_3^2\tau_4^2 + 394\tau_2^2\tau_4^3 + 
      216\tau_4^4 
\\&& - 22\tau_2^4\tau_3\tau_5 + 3\tau_2\tau_3^3\tau_5 
     - 448\tau_2^2\tau_3\tau_4\tau_5 - 
      390\tau_3\tau_4^2\tau_5 + 550\tau_2^3\tau_5^2 - 
      2325\tau_3^2\tau_5^2 
\\&& + 4225\tau_2\tau_4\tau_5^2 + 88\tau_2^5\tau_6 
     + 246\tau_2^2\tau_3^2\tau_6 + 444\tau_2^3\tau_4\tau_6 
     + 1683\tau_3^2\tau_4\tau_6 - 5020\tau_2\tau_4^2\tau_6 
\\&&  + 11220\tau_2\tau_3\tau_5\tau_6 + 22125\tau_5^2\tau_6 
     - 10854\tau_2^2\tau_6^2 + 8730\tau_4\tau_6^2 ]  
\\[1mm]&&
  + 8 \pi^4[42\tau_2^4\tau_3^2 - 72\tau_2\tau_3^4 - 112\tau_2^5\tau_4 + 
  832\tau_2^2\tau_3^2\tau_4 - 1804\tau_2^3\tau_4^2 + 179\tau_3^2\tau_4^2 
\\&& 
  - 1696\tau_2\tau_4^3 + 236\tau_2^3\tau_3\tau_5 + 
  2943\tau_3^3\tau_5 - 3809\tau_2\tau_3\tau_4\tau_5 - 8435\tau_2^2\tau_5^2 
\\&&
  - 13075\tau_4\tau_5^2 - 1952\tau_2^4\tau_6 - 10899\tau_2\tau_3^2\tau_6 + 
   13288\tau_2^2\tau_4\tau_6 + 5008\tau_4^2\tau_6  
\\&& 
  - 46035\tau_3\tau_5\tau_6 + 38916\tau_2\tau_6^2] 
\\[1mm]&&
  + 16 \pi^6[34\tau_2^6 - 480\tau_2^3\tau_3^2 - 
  597\tau_3^4 + 1790\tau_2^4\tau_4 + 424\tau_2\tau_3^2\tau_4 + 
  4665\tau_2^2\tau_4^2 + 1182\tau_4^3 +
\\&& 
  3281\tau_2^2\tau_3\tau_5 + 9435\tau_3\tau_4\tau_5 + 
  17425\tau_2\tau_5^2 + 2692\tau_2^3\tau_6 + 
  17979\tau_3^2\tau_6 
\\&& - 29530\tau_2\tau_4\tau_6 - 25185\tau_6^2] 
\\[1mm]&&
  + 32 \pi^8[-540\tau_2^5 + 291\tau_2^2\tau_3^2 - 6338\tau_2^3\tau_4 - 
  2488\tau_3^2\tau_4 - 4222\tau_2\tau_4^2 - 9908\tau_2\tau_3\tau_5  
\\&& - 9775\tau_5^2 + 698\tau_2^2\tau_6 + 18322\tau_4\tau_6]  
\\[1mm]&& 
  + 64 \pi^{10}[2308\tau_2^4 + 1170\tau_2\tau_3^2 + 9556\tau_2^2\tau_4 + 
  1012\tau_4^2 + 6345\tau_3\tau_5 - 1980\tau_2\tau_6] 
\\[1mm]&& 
 - 16 \pi^{12}[35992\tau_2^3 + 7683\tau_3^2 + 53924\tau_2\tau_4 +
  372\tau_6]  
\\[1mm]&&
  + 32 \pi^{14}[36053\tau_2^2 + 15241\tau_4] - 1148928 \,\pi^{16}\tau_2 + 
  440128\,\pi^{18}\;.
\end{eqnarray*}
%%%%%%%%%%%%%%%%%%%%%%%%%%%%%%%%%%%%%%%%%%%%%%%%%%%%%%%%%%%%%%%%%%%%

%\include{bibl}

%%%%%%%%%%%%%%%%%%%%%%%%%%%%%%%%%%%%%%%%%%%%%%%%%%%%%%%%%%%%%%%%%%%

\end{document}